\documentclass[aps,twocolumn,floatfix,
superscriptaddress]{revtex4}
\usepackage{amssymb,amsmath,graphicx,times}
\usepackage{mathptmx}
\usepackage[usenames]{color}

\usepackage{amsfonts,amsthm} 
\usepackage{bm,bbm}
\usepackage{physics}
\usepackage{mathtools}
\usepackage{lscape}
\usepackage{multirow}
\usepackage{diagbox}
\usepackage{color}
\usepackage[dvipsnames]{xcolor}
\usepackage{hyperref}
\usepackage{tensor}

\usepackage{soul}

\def\ket#1{{|#1\rangle}}
\def\bra#1{{\langle#1|}}

\usepackage{enumitem}

\newcommand{\vc}{\vcentcolon =}
\newcommand{\cv}{= \vcentcolon}

\newcommand{\ordW}{p}	
\newcommand{\Wcl}{W^{cl}}		
\newcommand{\Wq}{W}		
\newcommand{\h}{\mathcal{H}}	
\newcommand{\Sh}{\h_S}	
\newcommand{\Ah}{\h_A}	
\newcommand{\bone}{\mathbbm{1}}	
\newcommand{\fs}{F_S}

\DeclareMathOperator{\diag}{diag}

\begin{document}
\title{ Quantum Monge--Kantorovich problem
   and transport distance between density matrices}
        
\author{Shmuel Friedland}
\affiliation{Department of Mathematics, Statistics and Computer Science, University of Illinois, Chicago, IL 60607-7045, USA}

\author{Micha{\l} Eckstein} 
\affiliation{Institute of Theoretical Physics, 
Jagiellonian University, ul. {\L}ojasiewicza 11, 30--348 Krak\'ow, Poland}

\author{Sam Cole}
\affiliation{Department of Mathematics, University of Missouri, Columbia, MO 65211, USA}
 
\author{Karol {\.Z}yczkowski}
\affiliation{Institute of Theoretical Physics, 
Jagiellonian University, ul. {\L}ojasiewicza 11, 30--348 Krak\'ow, Poland}
\affiliation{Center for Theoretical Physics, Polish Academy of Sciences, Al. Lotnik\'{o}w 32/46, 02-668 Warszawa, Poland}


\date{September 17, 2021}

\begin{abstract}
\noindent

A quantum version of the Monge--Kantorovich optimal transport problem is analyzed.
The transport cost is minimized over the set of all bipartite coupling  states
 $\rho^{AB}$,
 such that both of its reduced density matrices $\rho^A$ and $\rho^B$ of dimension $N$
 are fixed.
We show that, selecting the quantum cost matrix to be proportional to the projector
on the antisymmetric subspace, the minimal transport cost
leads to a semidistance between $\rho^A$ and $\rho^B$,
which is bounded from below by the rescaled Bures distance and from above by the root infidelity.
In the single qubit case we provide a semi-analytic 
expression for the optimal transport cost between any two states 
and prove that its square root satisfies the triangle inequality 
and yields an analogue of the Wasserstein distance of order two on the set of density matrices.
We introduce an associated measure of proximity of quantum states,
called  SWAP-fidelity,  and discuss its properties and applications in quantum machine learning.

\end{abstract}

\maketitle

\emph{Introduction.}---Remarkable progress in quantum technologies 
 stimulates further research on foundations of quantum mechanics. In particular, 
one aims to improve our understanding of the structure 
of the set of quantum states \cite{BZ17} -- the arena in which 
quantum information processing takes place.
It is therefore important to analyze various distances in the space
of quantum states and to describe their properties and 
diverse physical applications.

In the classical case one considers several distances in the space of probability distributions.
A prominent role is played by the Monge distance, directly 
linked to the famous mass transport problem \cite{Mo1781},
in which one minimizes the work against friction required 
to move a pile of earth of shape $p^A$ into the final shape $p^B$.
For continuous distributions the problem is solved analytically 
in any 1D case \cite{Sal43} and 
in several particular 2D cases \cite{RR98},
while effective numerical algorithms can be applied
for any discrete probability distributions.

More general formulations of Kantorovich \cite{Kan42,Kan48} 
and Wasserstein \cite{Was69},
relying on joint probability distributions with marginals $p^A$ and $p^B$,
are explicitly symmetric with respect to given probability distributions.
Due to numerous applications of the mass transport problem
 in operations research and economics
and its relation to the assignment problem,
it remains a subject of intensive mathematical research \cite{Vi09,Ve13}.
The transport problem was inspected from 
the perspective of free probability \cite{BV01} and applied in the study of causality \cite{EM17a,EM17b,EHMH20}.
 
An attempt to generalize the notion of the Monge distance
for quantum theory was pursued for the setup of infinite \cite{ZS98}
and finite dimensional \cite{ZS01,BZ17} Hilbert spaces.
Such a  distance between any two quantum states, defined
by the Monge distance between the corresponding Husimi distributions,
enjoys the semiclassical property: the distance between two coherent states,
centered at points $x$ and $y$  in the classical phase space,
is equal to the distance, $|x-y|$, between the points
at which both coherent states are concentrated. This property,
crucial for studies on quantum analogues of the Lyapunov exponent  \cite{ZWS93},
is also shared by the distance recently proposed in \cite{WWW20}.

Any definition based on the notion of the Husimi function 
depends on the choice of the set of coherent states.
It is therefore natural to look for a universal method to introduce 
the transport distance between quantum states 
directly by applying the Kantorovich--Wasserstein approach
and performing optimization over the set of bipartite
quantum states with fixed marginals \cite{AF17,GP18,CGT18}.
In spite of recent  vibrant activity in this field 
\cite{GMT16,CGNT17,BGJ19,FGZ19,CGGT17,Fri20,Du20b},
 this aim has not been fully achieved until now \cite{YZYY18,Reira18,Ikeda20,PMTL20}. 
 
In parallel, the quantum optimal transport has found numerous applications in quantum physics, in particular in connection with the measures of proximity of quantum states \cite{DPT19,CM20,DR20,CGP20}. The latter play a key role in quantum metrology \cite{BC94,Sa17,LYLW20}, as well as in quantum machine learning \cite{QML,LW18,
Wu19,KdPMLL21}.
\smallskip

In this work we introduce  a measure of proximity of quantum states,
dubbed the ``SWAP-fidelity'', as it is inspired by the quantum optimal transport with a specific quantum cost matrix. It shares many properties with the standard Uhlmann--Jozsa  fidelity \cite{Uh76,Jo94} and agrees with the latter if at least one of the states is pure. We prove that the square root of the associated quantum optimal transport cost yields 
a new distance on the space of qubits which is a quantum analogue of the Wasserstein-2 distance. For larger dimensions we show analytically that this quantity gives a semidistance, bounded from above by the root infidelity
and from below by the rescaled  Bures distance. We further discuss the general form of a quantum cost matrix and study the quantum-to-classical transition of the transport problem. The latter shows that the quantum optimal transport is cheaper than the classical one, generalizing the results of \cite{CGP20}. Finally, we discuss an application of the new quantum metric in the context of quantum generative adversarial networks.

 
\medskip 
\emph{Classical  transport problem.}---To formulate the mass transport problem
 in the setup of Kantorovich for any probability distributions $p^A_i$ and $p^B_j$ 
 one introduces the notion of a classical coupling  
  --  a joint probability distribution $P^{AB}_{ij}$ with two specified marginals, 
  $p^A_i=\sum_j P^{AB}_{ij}$ and  $p^B_j=\sum_i P^{AB}_{ij}$. 
  In the case of two
   probability vectors of order $N$,
  $p^A, p^B\in \Delta_N$,
 any joint distribution $P^{AB}\in \Delta_{N^2}$,
 which determines a transport plan,
 is represented by a single vector  $P_\mu$,
   with $\mu=(i-1)N+j=1,\dots, N^2$.
 The set  $\Gamma^{cl}(p^A, p^B)$ of all admissible couplings
 forms a convex subset of the simplex $\Delta_{N^2}$, 
 with extreme points characterized in \cite{Pa07}.
 
Consider a  set of $N$  points $X \vc \{x_i\}_{i=1}^N$ equipped with a distance function $d$. With the latter we associate 
a symmetric $N \times N$ matrix $E_{ij} \vc d(x_i,x_j)$.
Assuming that the transport cost of a unit of mass from point $x_i$ to point $x_j$
is equal to the distance $E_{ij}$, 
one can formulate the classical transport problem \cite{RR98,Fri20}.
In order to study its quantum analogue it will be convenient
to reshape the square distance matrix $E_{ij}$ of dimension $N$ 
into a distance vector $D_{\mu}$ of length $N^2$.

  To generate a Wasserstein distance
   between probability distributions $p^A$ and $p^B$
   one can use a $N^2 \times N^2$ classical cost matrix $C$
which is diagonal,
 $C_{\mu\nu}=D_{\mu} \delta_{\mu\nu}$ for $\mu,\nu=1,\dots, N^2$,
 and a diagonal density matrix $\rho^{AB}_{\mu\nu}=P_{\mu} \delta_{\mu\nu}$.
For a given transport plan  $P^{AB}$ 
and some parameter $p \geq 1$ the total transport cost is given
 by the scalar product, 
 ${\hat T}_{C,\ordW}(P) \vc \sum_{\mu=1}^{N^2} (D_\mu)^\ordW P_\mu = \Tr\,C^\ordW \rho^{AB}$.
  The minimal transport cost,  $T_{C,\ordW}^{cl} \vc \min_P {\hat T}_{C,\ordW}(P)$, optimized for a given value
  of the parameter $\ordW$, leads to the family of Wasserstein distances,
        \begin{equation}
      \!\!\Wcl_{C,\ordW}(p^A,p^B) \vc
      \Bigl( \min_{P^{AB}\in \Gamma^{cl}} ( \Tr\, C^{\ordW} \rho^{AB} ) \Bigr)^{\!1/\ordW}\! = \big(T_{C,\ordW}^{cl} \big)^{1/\ordW}. 
      \label{class_opt}
      \end{equation}
The minimum is  taken over the set  $\Gamma^{cl}(p^A,p^B)$ of classical couplings \cite{RR98}.
If $d(x_i,x_j) = 1-\delta_{ij}$, the space $X$ has
the geometry of an $N$-point simplex $\Delta_N$. 
In this case, $C^p = C$ and $W^{cl}_{C,p} = (W^{cl}_{C,1})^{1/p}$ for any $p \geq 1$, so we shall abbreviate $T^{cl} \vc T_{C,p}^{cl}$ and denote the classical cost matrix by $C^{cl}$.

\medskip
\emph{Proximity of quantum states.}---We now switch to the quantum setting and denote the set of $N \times N$ density matrices by  
       $\Omega_N=\{\rho: \ \rho=\rho^{\dagger}, \ \rho\ge 0, \ \Tr \rho=1\}$. To quantify the closeness of any two quantum states one uses various distances on $\Omega_N$ -- see \cite{BZ17}.
The {\sl trace distance}, singled out by the Helstrom theorem on optimal distinguishability \cite{Helstrom}, reads
$D_{\rm Tr} (\rho^A, \rho^B) \vc \frac{1}{2} {\rm{Tr}}|\rho^A - \rho^B|$,
   where $|X|:=\sqrt{XX^{\dagger}}$. Another way to characterise the proximity between two density matrices relies on Uhlmann--Jozsa {\sl fidelity} \cite{Uh76,Jo94},
$F(\rho^A, \rho^B) := \big({\rm {Tr}} |\sqrt{\rho^A} \sqrt{\rho^B}|  \big)^2$. It leads to the following distances: the  {\sl root infidelity}  \cite{GLN05}, 
 $I \vc \sqrt{1-F}$, the {\sl  Bures distance } \cite{Uh76,Uh95},
   $B \vc \sqrt{2(1-\sqrt{F})}$, and the {\sl Bures angle}, $A \vc \tfrac{2}{\pi} \arccos \sqrt{F}$. 
   Note that the Bures distance and other distances based on fidelity 
   are closely related to statistical distinguishability 
   and quantum Fisher information \cite{BC94,Sa17},
   so they have a direct interpretation in quantum metrology \cite{LYLW20}.
   
   We shall now introduce a quantity analogous to fidelity, which is directly related to the quantum optimal transport problem and its applications in machine learning \cite{Wu19}.

\medskip 
\emph{SWAP-fidelity.}---  Consider two arbitrary states $\rho^A, \rho^B \in \Omega_N$.
A composed (bipartite) density matrix  $\rho^{AB}$ of order $N^2$
 is called a {\sl coupling matrix} \cite{Wi15} between $\rho^A$ and $\rho^B$ 
 if both  partial traces agree, $\Tr_A \rho^{AB}=\rho^B$ and 
               $\Tr_B \rho^{AB}=\rho^A$.
          The set of all possible quantum couplings matrices 
           will be denoted by     
   $\Gamma^{Q}(\rho^A, \rho^B) \subset \Omega_{N^2}$ -- see Fig.1c.  The bipartite quantum states can be conveniently represented in the 
{\sl Fano form} -- see Supplemental Material (SM).
\begin{figure}[h]
	\centering
	\includegraphics[width=1\linewidth]{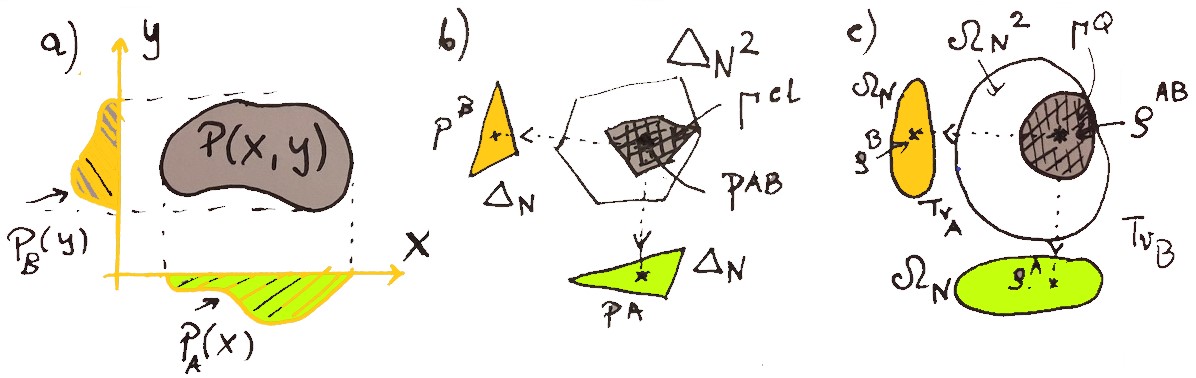}
	 \caption{Couplings between probability distributions used for Kantorovich distance: a) continuous 1D probabilities $p_A(x)$ and $p_B(y)$ coupled by
	 a  joint distribution $P(x,y)$;
	  b) two  $N$-point classical states  $p^A,p^B\in \Delta_N$ coupled
       by a joint state $P^{AB}\in \Gamma^{cl} \subset \Delta_{N^2}$ with adjusted marginals;
       c) two quantum states $\rho^A, \rho^B \in \Omega_N$ coupled by a 
           bipartite state  $\rho^{AB}\in \Gamma^Q\subset \Omega_{N^2}$
             such that  $\Tr_A \rho^{AB}=\rho^B$ and 
               $\Tr_B \rho^{AB}=\rho^A$.
}
        \label{fig1}
\end{figure}

Let $S$ denote the SWAP operator,
$S(|x\rangle \otimes |y\rangle) := |y\rangle \otimes |x\rangle$
for any vectors $|x\rangle$ and $|y\rangle $. For any $\rho^A, \rho^B \in \Omega_N$ we introduce the \emph{SWAP-fidelity}:
\begin{equation}
\fs(\rho^A, \rho^B) \vc
 \max_{\rho^{AB}\in \Gamma^Q} \big( \Tr\, S \, \rho^{AB} \big),
\label{fS}
\end{equation}
where the maximum is taken over the set  $\Gamma^Q$ of all admissible coupling matrices 
$\rho^{AB}$.

 
\smallskip \noindent
{\bf Proposition 1}. {\sl For any dimension $N$, the SWAP-fidelity $\fs$ is a symmetric jointly concave function from $\Omega_N \times \Omega_N$ to the unit interval. Furthermore, $\fs(\rho^A, \rho^B) = 1$ iff $\rho^A = \rho^B$, $\fs(\rho^A, \rho^B) = 0$ iff \,$\Tr \rho^A \rho^B = 0$, and 
\begin{align}
& \fs \bigl(\rho^A, \rho^B \bigr)  =  \fs \bigl( U\rho^AU^{\dagger}, U\rho^BU^{\dagger}\bigr),  \text{ for } \, U \in \mathrm U(N), \label{fSU} \\
& F(\rho^A, \rho^B) \leq \fs(\rho^A, \rho^B) \leq \sqrt{F(\rho^A, \rho^B)}, \label{fSB} \\
& \fs \bigl(\rho^A \otimes \sigma^A, \rho^B \otimes \sigma^B \bigr) \geq \fs \bigl(\rho^A, \rho^B\bigr) \fs \bigl(\sigma^A, \sigma^B \bigr). \label{fSM}
\end{align}
}
%

The above result, proven in SM, shows that, in analogy to the fidelity $F$, the SWAP-fidelity $\fs$ equals unity iff both states coincide and vanishes iff they are orthogonal. Furthermore, it interpolates between fidelity and root fidelity -- see Ineq. \eqref{fSB} -- with the first inequality saturated if at least one of the states is pure. Notably, the SWAP-fidelity is {\sl super-multiplicative} with respect to tensor product 
as it satisfies inequality  \eqref{fSM} characteristic to superfidelity \cite{MPHUZ}.
Note also that $\fs$ is jointly concave, as is the root fidelity $\sqrt{F}$, while it has a probabilistic interpretation for pure states, $\fs (\phi,\psi) = F(\phi,\psi)  = \vert \langle \phi , \psi \rangle \vert^2$ -- see \cite{BZ17}. 
The SWAP-fidelity is shown below to be closely related to the quantum optimal transport and yields a novel metric on the Bloch ball.

\medskip 
\emph{Quantum cost matrix.}---
To study the
transport problem between two quantum states of order $N$
we need to specify a quantum cost matrix $C^Q$ of size $N^2$. 
Let $\{\ket{i}\}_{i=1}^N$ be the computational basis of an $N$-level quantum system and denote the maximally entangled singlet states 
in the subspace spanned by $|i\rangle $ and $|j\rangle$
by $|\psi^-_{ij}\rangle=\frac{1}{\sqrt{2}}(|i,j\rangle -|j,i\rangle)$.
In the case of the simplex geometry,   $E_{ij} = 1-\delta_{ij}$,
the quantum optimal transport enjoys several
 desirable features if one chooses
 the cost matrix $C^Q$ to be the projector onto the antisymmetric subspace,
\begin{equation}
 C^Q  = \sum_{j>i=1}^{N} |\psi^-_{ij}\rangle  \langle \psi^-_{ij}| = \tfrac{1}{2} ({\mathbbm 1}_{N^2}- S) = (C^Q)^2,
      \label{costLn}
      \end{equation}
   as advocated also in \cite {YZYY18,Reira18,Wu19,Du20b}.
In particular, for the simplest, one-qubit problem, $N=2$,  
the cost matrix reads         
\begin{equation}
C^Q  = \frac{1}{2} 
  \left[
\begin{array}{llll}
   0  & \ 0 & \ 0 & 0  \\
  0  & \ 1 & \!\!\!-1 & 0  \\
   0  & \!\!\!-1 & \ 1 & 0  \\
    0  &\ 0 & \ 0 & 0  
\end{array}
\right] = |\psi^-\rangle  \langle \psi^-|
\label{cost_alpha}
\end{equation}
The above quantum cost matrix $C^Q$  of size $N^2$
forms a {\sl coherification} \cite{KCPZ18}
of a classical cost matrix $C^{cl}={\rm diag}(C^Q)$ corresponding to the simplex geometry.
   
We are now going to look for the minimal quantum transport cost, which can be expressed using the SWAP-fidelity:
\begin{equation}
T^{Q}(\rho^A, \rho^B) \vc
 \min_{\rho^{AB}\in \Gamma^Q} \big( \Tr\, C^Q \rho^{AB} \big) = \frac{1 - \fs(\rho^A,\rho^B)}{2}.
\label{cost}
\end{equation}

Proposition 1 directly implies that for any two states $\rho^A, \rho^B \in \Omega_N$
the optimal quantum transport cost, $T^Q$, is jointly convex, symmetric,
non-negative and vanishes iff $\rho^A = \rho^B$.
 Furthermore, for $C^Q$ given by \eqref{costLn} one has
\begin{equation}
T^Q\bigl(\rho^A, \rho^B \bigr)  =  T^Q\bigl( U\rho^AU^{\dagger}, U\rho^BU^{\dagger}\bigr),
  \label{T_invariance}
      \end{equation}
for any unitary operator $U$ on $\mathbb{C}^N$. Hence, $T^Q$ forms a semidistance on $\Omega_N$, as shown independently in \cite{Wu19}.
In analogy with the classical definition \eqref{class_opt}, for any $p \geq 0$ we introduce a quantum analogue of the $p$-Wasserstein distance, $W_p \vc (T^Q)^{1/p}$. As shown below,  $\Wq_2$ plays a distinguished role, so we will denote it simply by $W \vc W_2$.

As an immediate corollary of Ineq. \eqref{fSB}, proven in SM with help of recent results by Yu et al.  \cite{YZYY18},  we arrive at explicit bounds for the quantum transport cost and its square root $W$.
%
%
%

\smallskip
{\bf Corollary 2}. {\sl For any two states $\rho^A, \rho^B \in \Omega_N$
we have
\begin{equation}
\!\!\!\!\!\tfrac{1}{\sqrt{2}} B(\rho^A,\rho^B) \ge  \tfrac{1}{\sqrt{2}} I(\rho^A,\rho^B) \ge  \Wq(\rho^A,\rho^B)  \ge
 \tfrac{1}{2}   B(\rho^A,\rho^B) .
      \label{T_bounds}
      \end{equation}
Furthermore, the second inequality is saturated if either $\rho^A$ or $\rho^B$ is pure.
}

\medskip 
\emph{Single qubit transport.}---In the simplest case of $N=2$
 the quantum cost matrix is given by 
 (\ref{cost_alpha}). 
 As shown in SM
the full solution of the quantum transport problem
for the one-qubit case is equivalent to solving a polynomial equation of order six. The latter yields analytic expressions in several special cases.
 
For two diagonal matrices,
$\rho^{cl}_r={\rm \diag}(r,1-r)$ and $\rho^{cl}_s={\rm \diag}(s,1-s)$,
 we have 
\begin{equation}
\Wq \big(\rho^{cl}_r, \; \rho^{cl}_s\big)
= \tfrac{1}{\sqrt{2}} \max\big\{ \big\vert \sqrt{r} - \sqrt{s} \big\vert, \big\vert \sqrt{1-r} - \sqrt{1-s} \big\vert \big\}.
  \label{T_diag}
      \end{equation}
Furthermore, if one of the states is totally mixed we obtain,
\begin{equation}\label{T_mixed}
\Wq \bigl(\tfrac{1}{2} {\mathbbm 1}, \rho \bigr) = \tfrac{1}{2} \max \big\{ \big\vert 1 - \sqrt{2\lambda} \big\vert, \big\vert 1 - \sqrt{2(1-\lambda)} \big\vert \big\},
\end{equation}
where 
$\lambda$ and $1-\lambda$ denote the eigenvalues of $\rho$.
A surprisingly simple formula is available for two isospectral states,  
\begin{equation}
\Wq \bigl(\rho, \;U\rho U^{\dagger} \bigr) = \sqrt{\tfrac{1}{\sqrt{2}}-\sqrt{\lambda(1-\lambda)}} \; \big\vert \sin (\theta/2) \big\vert,
  \label{T_isosp}
      \end{equation}
where $\{\lambda, 1-\lambda\}$ is the common spectrum and $\theta$ is the $U$-dependent angle between Bloch vectors associated with 
the states --  see SM.
In the case of  a single qubit we obtain one of the key results of this work,
proved in SM.
\medskip

{\bf Theorem 3}. {\sl For $N=2$
the function $\Wq_p$
 satisfies the triangle inequality iff $p \geq 2$: 
for any states $\rho^A, \rho^B, \rho^C \in \Omega_2$
one has
$\Wq_p(\rho^A, \rho^B)+\Wq_p(\rho^B, \rho^C) \ge \Wq_p(\rho^A, \rho^C)$.
Thus, $W_p$ generates a distance on the Bloch ball $\Omega_2$,
analogous to the classical $p$-Wasserstein distance 
 (\ref{class_opt}), provided that $p \geq 2$.}

\smallskip

Numerical studies carried out for  the simplex geometry 
with $N=3$ and $4$  
allow us to conjecture that $W_p$ is actually 
a distance on $N$-level systems for any $N$ and $p \geq2$.
Whereas the quantum transport cost itself, $T^Q = W_1$, is \emph{not} a distance on $\Omega_2$, its square root, $W_2$, is -- see SM. 
Note that, similarly, while the infidelity $1-F$ does not satisfy the triangle inequality, its square root, $I = \sqrt{1-F}$, does \cite{GLN05}.
An analogous property was recently proved for the quantum Jensen--Shannon divergence, the square root of which 
satifies the triangle inequality and
leads to the transmission metric \cite{BH09,Vi19}.

\begin{figure}[h]
	\centering
		\includegraphics[width=0.9\linewidth]{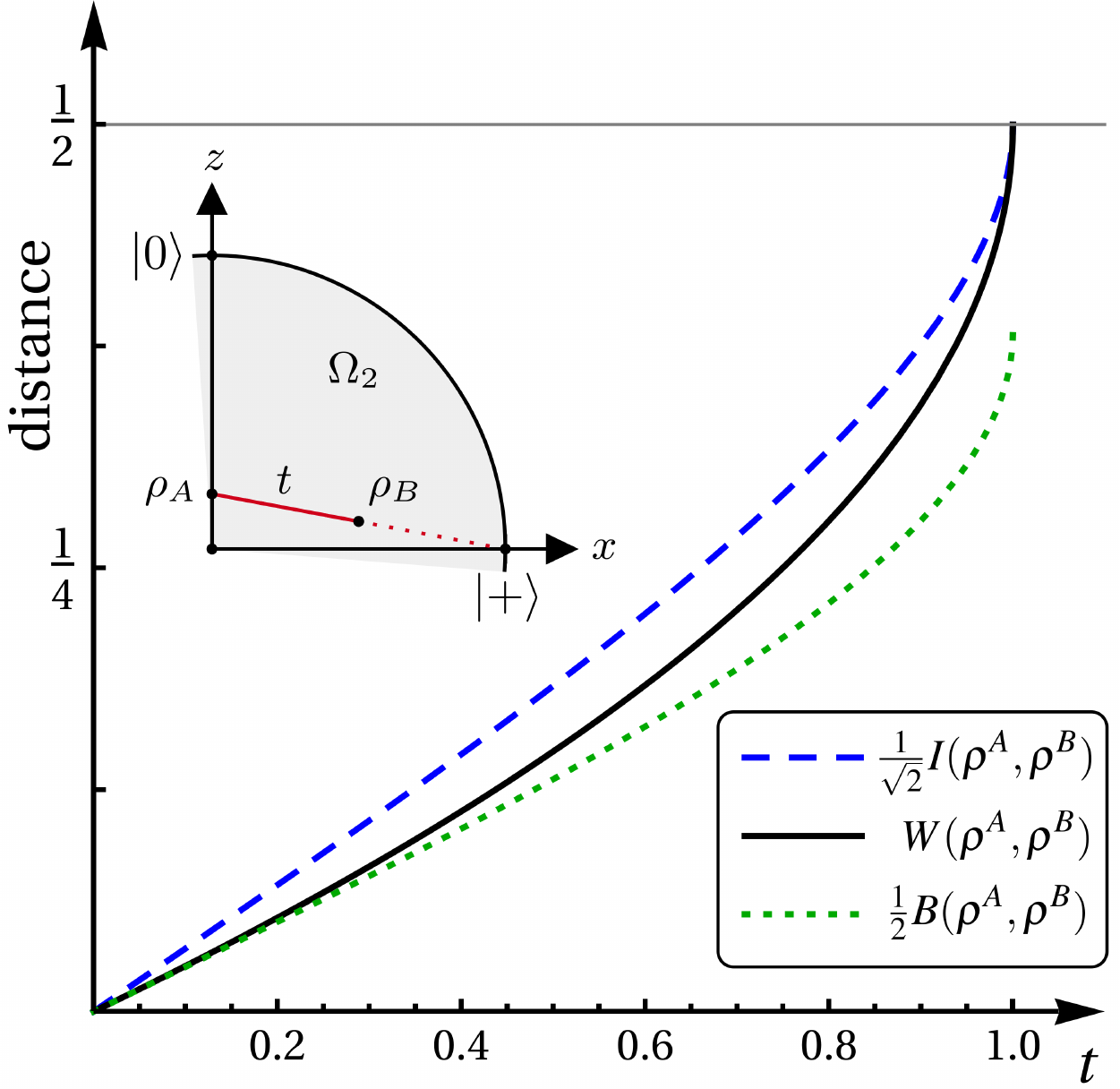}
	 \caption{Bounds (\ref{T_bounds}) 	 illustrated 
	  in the Bloch ball $\Omega_2$.
	 The distances 
	   between the state $\rho^A = \tfrac{9}{20} {\mathbbm 1} + \tfrac{1}{10} \ket{0}\bra{0}$ and $\rho^B = (1-t) \rho^A + t (\ket{+} \bra{+})$
	   are shown as a function of $t \in [0,1]$ varying along the Euclidean line (red line in the inset).
}
        \label{fig2}
\end{figure}

 Recall also that the Monge distance between quantum states
 defined by the Husimi distribution with respect to spin coherent states
  for $N=2$ leads to the Hilbert--Schmidt distance 
 and the Euclidean geometry on the Bloch ball,
 while the discrete Monge distance, describing movements of the
 Majorana stars corresponding to pure states, 
 gives geodesic distance on the Bloch sphere \cite{ZS01}.
Whereas formula \eqref{T_bounds} implies that $W$ is strongly equivalent to the Bures metric $B$ and induces the same topology on the Bloch ball, the corresponding (curved) geometries are different. This is illustrated in Fig. \ref{fig2} for a fixed mixed state $\rho^A$ and $\rho^B$ varying continuously from $\rho^A$ to the pure state $\ket{+}$. We witness the validity of the bound \eqref{T_bounds}, with $W(\rho^A,\ket{+}) = I (\rho^A,\ket{+})$. Observe also that initially the transport distance curve closely follows that of the Bures distance. Using the Pauli matrices, $\sigma_i$, and the Bloch representation, $\rho_\pm(\vec{\tau}) \vc \tfrac{1}{2} \left( {\mathbbm 1} \pm \vec{\tau} \cdot \vec{\sigma} \right)$ for $\Vert \vec{\tau} \Vert \in [0,1]$, we have
\begin{align}\label{WB}
\Wq \bigl(\rho_+(\vec{\tau}),\rho_-(\vec{\tau}\bigr) = 
\tfrac{1}{\sqrt{2}} B \bigl(\rho_+(\vec{\tau}),\rho_-(\vec{\tau})\bigr). 
\end{align}

\begin{figure}[h]
	\centering
	\includegraphics[width=0.75\linewidth]{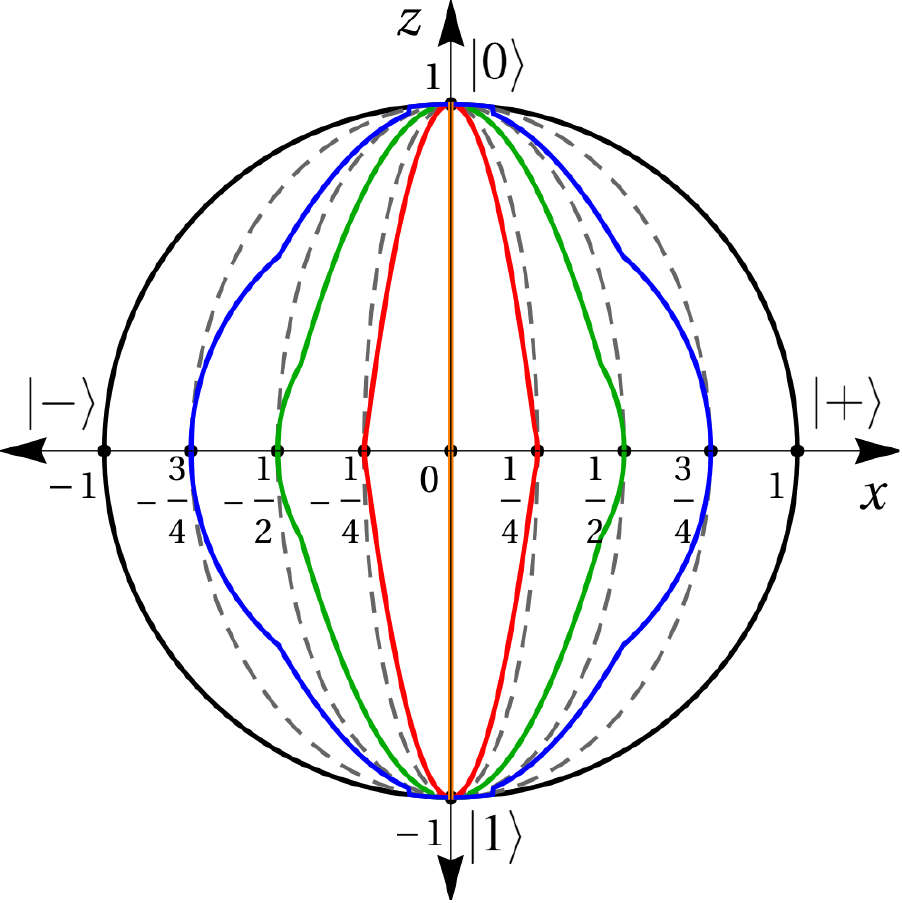}
	 \caption{The $y=0$ section of the Bloch ball $\Omega_2$. Solid colored curves represent the geodesics of the transport metric $\Wq$, while dashed lines are ellipses corresponding to the geodesics of the Bures angle $A$.
}
        \label{fig3}
\end{figure}

Another insight into the geometry induced by the transport distance $W$ is gained by the study of {\sl geodesics} -- trajectories in the space of states -- on which the triangle inequality is saturated for every triple of points. Such geodesics do not exist for either the root infidelity or for the Bures distance. On the other hand, the geodesics of the Bures angle metric $A$ exist and have a nice geometrical interpretation \cite{Uhlmann} as great circles on the Uhlmann 3-hemisphere $\tfrac{1}{2}S^3$. When projected onto the equatorial plane they form ellipses within the real (i.e. $y=0$) slice of the Bloch ball. Interestingly, such geodesics also exist for the transport metric $W$, though their shape is more complex, as shown in Fig. \ref{fig3}.

\medskip 
\emph{Quantization of an arbitrary classical cost matrix.}---In
 a more general set-up  consider an arbitrary distance function 
 on the $N$-point set $X$, determined by the matrix  
$E_{ij}$.
 With any such classical geometry of $X$ we  associate 
 the following quantum cost matrix 
\begin{equation}
  C^Q_E = \sum_{j>i=1}^{N} E_{ij} |\psi^-_{ij}\rangle \langle \psi^-_{ij}|.
      \label{costWnbis}
      \end{equation} 
Accordingly, for any $p \geq 1$ we define
\begin{align*}
T^{Q}_{E,p}(\rho^A, \rho^B) \vc \min_{\rho^{AB}\in \Gamma^Q} \big( \Tr\, (C^Q_E)^p \rho^{AB} \big), && W_{E,p} \vc \big(T^{Q}_{E,p}\big)^{1/p}
\end{align*}
and prove
the following result in SM.
\medskip

{\bf Proposition 4.} 
 {\sl For any $N$, any $p \geq 1$, and any choice of classical geometry $E$, the
  $p$\textsuperscript{th} root of the corresponding optimal quantum transport cost,
   $W_{E,p}$,  is a semidistance on $\Omega_N$.}

\medskip 
\emph{The quantum-to-classical transition.}---It is instructive to compare the quantum transport problem with its classical counterpart. To this end one embeds classical probability vectors in diagonal density matrices. The following result shows that the quantum transport cost between two classical states is always cheaper than the corresponding classical cost (see also \cite{CGP20}).

\smallskip

{\bf Proposition 5.}
{\sl Let $\vec{r},\vec{s}$ be two $N$-dimensional probability vectors and let $\rho^{cl}_{\vec{r}}, \rho^{cl}_{\vec{s}} \in \Omega_N$ be the corresponding quantum states defined as $(\rho^{cl}_{\vec{r}})_{ij} \vc r_i \delta_{ij}$.} Then, 
\begin{align*}
     T^Q \bigl(\rho^{cl}_{\vec{r}},  \rho^{cl}_{\vec{s}} \bigr) \ \le \ 
      T^{cl}\bigl(\vec{r},  \vec{s} \bigr).
\end{align*}
This follows from the fact that the quantum optimization is performed over a strictly larger set of admissible couplings, $\Gamma^{cl}(\vec{r}, \vec{s}) \subset \Gamma^{Q}(\rho^{cl}_{\vec{r}},  \rho^{cl}_{\vec{s}})$.

The quantum-to-classical transition of the transport problem can be interpreted in terms of decoherence caused by the interaction of the information processing device with its environment. As a simple model (cf. \cite{Zurek03}) one can assume that the quantum cost matrix is acted upon by a dephasing channel $\mathcal{E}_{\alpha}(C^Q) = \alpha C^Q + (1-\alpha) \diag (C^Q)$, with the parameter $\alpha \in [0,1]$ proportional to the $l_1$-coherence \cite{BCP14}. One can then study the function $T^Q_{\alpha} \vc \min_{\rho^{AB}\in \Gamma^Q} \big( \Tr\, \mathcal{E}_{\alpha}(C^Q) \rho^{AB} \big)$. In the single-qubit case it is easy to check (see SM) that $\sqrt{T^Q_{\alpha}}$ is a distance on the set of commuting density matrices of order 2. Moreover, it is a strictly decreasing function of $\alpha$, provided that the two input states are different and none of them is pure.

\medskip 
\emph{Applications.}---The introduced SWAP-fidelity offers an original measure of proximity between quantum states and thus provides a new tool to quantify 
protocols of quantum information processing. Its most promising and straightforward application pertains to quantum generative adversarial networks (QGANs) \cite{LW18,DdK18}. This protocol of quantum machine learning \cite{QML} consists of a generator, which produces ``fake'' data, and a discriminator, which aims at distinguishing between the real and fake input data. The adversarial training reaches a fixed point when the generator produces data with true statistics and the discriminator's efficiency is 50\%. Similar to with classical GANs, the choice of the distance between real and fake data is critical for the stability and performance of the training \cite{WGAN,Wu19,LYLW20}. In \cite{LYLW20} it was argued that problems with efficiency of quantum learning algorithms \cite{PT11,MBSBN18,WFCSSCC20,CSVCC21} arise because the employed measure of proximity diminishes exponentially with the number of qubits. Although the introduced SWAP-fidelity suffers from the same drawback for pure states (as it equals to fidelity in this case), it might prove superior for mixed states because of its super-multiplicativity \eqref{fSM}.

In fact, in \cite{Wu19} a QGAN based on the semidistance $T^Q$ was shown to exhibit improved performance over other QGANs. Furthermore, it is noise tolerant and can be successfully used to approximate complicated quantum circuits with a limited number of quantum gates. Our results suggest that the choice $W = \sqrt{T^Q}$ is superior to $T^Q$, as it forms a genuine distance. 


 \medskip 
\emph{Outlook and conclusions.}---
We studied the quantum transport problem for density matrices of dimension $N$ with a cost matrix $C^Q$ taken to be a projector onto the antisymmetric subspace.
In the case of any two single-qubit states
we presented a constructive procedure to compute the quantum transport cost. A more detailed mathematical study is presented in the companion paper \cite{CEFZ21}.

 Inspired by the Wasserstein distance of order $2$,
 we proved that the square root of the optimal transport cost,
 $W_2=\sqrt{T^Q}$,
yields a distance on the Bloch ball, bounded by
 the  rescaled Bures distance and the root infidelity.
In the general problem of $N$-level systems and arbitrary classical geometry $E$ we showed that an analog of the $p$-Wasserstein distance $W_p$ yields a semidistance on the full space of quantum states for any $p\ge 1$. Furthermore, numerical results allow us to conjecture that $\sqrt{W_1}$ enjoys the triangle inequality in full generality.

Given the multifarious applications of the classical Wasserstein distances, we expect its quantum analogue to play a pivotal role in diverse branches of quantum information processing. Furthermore, the   SWAP-fidelity -- a novel quantity introduced in this work -- is likely to offer new advances in characterizing proximity between quantum states.
\medskip

\emph{Acknowledgements.}---It is a pleasure to thank W.~S{\l}omczy{\'n}ski 
for inspiring discussions and helpful remarks. 
Financial support by Simons collaboration grant for mathematicians, Narodowe Centrum Nauki 
under the Maestro grant number DEC-2015/18/A/ST2/00274 
and by Foundation for Polish Science 
under the Team-Net project no. POIR.04.04.00-00-17C1/18-00
is gratefully acknowledged.

\bigskip

\appendix

%

\section{\label{app:gen}Properties of the quantum optimal transport}

In this section we provide some details about the general quantum optimal transport problem, along with the proof of Proposition 4 from this article. 
For a complete mathematically oriented presentation of the problem
 the reader is invited to consult the companion-article \cite{CEFZ21}.

Let $\h = \mathbb{C}^N \otimes \mathbb{C}^N$ for some $N \geq 2$. The SWAP operator $S$ is a linear operator on $\h$, which acts on product states as $S(|x\rangle \otimes |y\rangle) = |y\rangle \otimes |x\rangle \cv \ket{y,x}$. Since $S^2 = \bone_{N^2}$ and $S = S^{\dagger}$ its eigenvalues are $\pm 1$. The Hilbert space $\h$ thus admits an orthogonal decomposition $\h = \Sh \oplus \Ah$ into symmetric and antisymmetric subspaces $\Sh$ and $\Ah$, respectively. The former can be identified with the symmetric complex matrices of order $N^2$, 
while the latter with the skew-symmetric ones.

\medskip

{\bf Definition A1}. {\sl A $N^2 \times N^2$ complex matrix $C$
 is called a \emph{quantum cost matrix} 
 if it is positive semidefinite,
  and $C\,X = 0$ if and only if
 the support of $X$ equals to $\Sh$.
 }

\smallskip

Explicit examples of quantum cost matrices are provided by formula \eqref{costWnbis} in the main body of the article. 
Note also that if $C$ is a quantum cost matrix, then so is $C^p$ for any $p > 0$.

\medskip

{\bf Definition A2}. {\sl Let $C$ be a quantum cost matrix of dimension $N^2$. 
 The associated \emph{quantum transport cost} is a map $T^Q_C : \Omega_N \times \Omega_N \to \mathbb{R}$ defined as
\begin{align}
T^{Q}_C(\rho^A, \rho^B) \;\; \vc \min_{\rho^{AB}\in \Gamma^Q(\rho^A,\rho^B)} \big( \Tr\, C \,\rho^{AB} \big),
\label{cost_gen}
\end{align}
with the set of quantum couplings}
\begin{align*}
 \Gamma^Q(\rho^A,\rho^B) =  \{ \rho^{AB} \in \Omega_{N^2} \; \vert \;  \Tr_A \rho^{AB}=\rho^B, \Tr_B \rho^{AB}=\rho^A\}.
\end{align*}

{\bf Proposition A3}. {\sl For any quantum cost matrix $C$, $T^{Q}_C$ is a convex function on  $\Omega_N \times \Omega_N$.}
\begin{proof}\renewcommand{\qedsymbol}{}
Let $\rho^A,\rho^B,\sigma^A,\sigma^B \in \Omega_N$ and let $a \in [0,1]$. Assume that $\rho^{AB} \in \Gamma^Q(\rho^A,\rho^B)$ and $\sigma^{AB} \in \Gamma^Q(\sigma^A,\sigma^B)$ are the optimal quantum couplings, i.e.
\begin{align*}
T^{Q}_C(\rho^A, \rho^B) = \Tr \, C \, \rho^{AB}, && T^{Q}_C(\sigma^A, \sigma^B) = \Tr \, C \, \sigma^{AB}.
\end{align*}
Define, for any $a \in [0,1]$, $\tau^{AB} \vc a \rho^{AB} + (1-a) \sigma^{AB}$. Then, $\tau^{AB} \in \Gamma^Q \big(a\rho^{A} + (1-a) \sigma^{A}, a\rho^{B} + (1-a) \sigma^{B} \big)$ and
\begin{multline}
T^{Q}_C(a\rho^{A} + (1-a) \sigma^{A}, a\rho^{B} + (1-a) \sigma^{B}) \\
\leq \Tr \, C \, \tau^{AB} = a T^{Q}_C(\rho^A, \rho^B) + (1-a) T^{Q}_C(\sigma^A, \sigma^B). \tag*{$\Box$}
\end{multline}
\end{proof}

The following central result implies Proposition 4 from the main text, because $C^Q_E$ are quantum cost matrices, and if $T^{Q}_C$ is a semidistance then so is $(T^{Q}_C)^{1/p}$ for any $p\geq 0$.

\medskip

{\bf Theorem A4}. {\sl Let $C$ be a $N^2 \times N^2$ quantum cost matrix. 
Then $T^{Q}_C$ is a semidistance on the space of $N$-level quantum systems $\Omega_N$.}

\begin{proof}\renewcommand{\qedsymbol}{}
We need to show that, for any $\rho^A,\rho^B \in \Omega_N$,
\begin{enumerate}[label=(\textit{\roman*})]
\item $T^{Q}_C(\rho^A, \rho^B)  \geq 0$,
\item $T^{Q}_C(\rho^A, \rho^B)  = 0$ if and only iff $\rho^B = \rho^A$,
\item $T^{Q}_C(\rho^A, \rho^B)  = T^{Q}_C(\rho^B, \rho^A)$.
\end{enumerate} 

\textit{(i)} Note that because $C$ is positive semidefinite and any $\rho^{AB} \in \Gamma^Q$ is a state in $\Omega_{N^2}$, we have $\Tr \, C \, \rho^{AB} \geq 0$. Hence, $T_C^Q ( \rho^A,\rho^B) \geq 0$ for any $\rho^A,\rho^B \in \Omega_N$.

\textit{(ii)} Suppose first that $\rho^A = \rho^B = \rho$. Given its spectral decomposition $\rho = \sum_{i=1}^N \sqrt{\lambda_i} \ket{i} \bra{i}$ we take the purification of $\rho$:
\begin{align*}
\rho^{\text{pur}} \vc \sum_{i,j = 1}^{N} \sqrt{\lambda_i \lambda_j} \ket{i} \ket{i} \bra{j} \bra{j}.
\end{align*}
Clearly, $\rho^{\text{pur}} \in \Gamma^Q(\rho,\rho)$. Since $\ket{i} \ket{i} \in \Sh$ we have, by definition of the quantum cost matrix, $C \big( \ket{i} \ket{i} \big) = 0$. Consequently, $\Tr C \rho^{\text{pur}} = 0$ and $T^{Q}_C(\rho, \rho) = 0$.

Suppose now, conversely, that $T^{Q}_C(\rho^A, \rho^B) = 0$. This implies that $\Tr C \rho^{AB} = 0$ for some $\rho^{AB} \in \Gamma^{A}(\rho^A,\rho^B)$. Consider its spectral decomposition $\rho^{AB} = \sum_{i=1}^{N^2} \mu_i \ket{X_i} \bra{X_i}$, where $\ket{X_i}$ are vectors in $\h$. We have $ \bra{X_i} C \ket{X_i} = 0$ for any $i$, which means that $\ket{X_i} \in \Sh$. But this implies that $\Tr_A \rho^{AB} = \Tr_B \rho^{AB}$, and hence $\rho^A = \rho^B$.

\smallskip

\textit{(iii)} A general state in $\Gamma^Q(\rho^A,\rho^B) \subset \Omega_{N^2}$ can be decomposed in an orthonormal basis as follows
\begin{align}\label{rhoAB}
\rho^{AB} = \sum_{i,j,k,\ell} c_{ijk\ell} \ket{i}\ket{j}\bra{k}\bra{\ell}.
\end{align}
Consequently, we have
\begin{align*}
\rho^A & = \Tr_B \rho^{AB} = \sum_{i,j,k} c_{ijkj} \ket{i}\bra{k},\\
\rho^B & = \Tr_A \rho^{AB} = \sum_{j,k,\ell} c_{kjk\ell} \ket{j}\bra{\ell}.
\end{align*}
Now, the SWAP operator extends to the space $\Omega_{N^2}$ and acts on the basis vector as follows:
\begin{align*}
S \big( \ket{i}\ket{j}\bra{k}\bra{\ell} \big) = S \big( \ket{i}\ket{j} \big) \bra{k}\bra{\ell} = \ket{j}\ket{i}\bra{k}\bra{\ell}.
\end{align*}
Hence, we have
\begin{align*}
S \rho^{AB} S^{\dagger} = \sum_{i,j,k,\ell} c_{ijk\ell} \ket{j}\ket{i}\bra{\ell}\bra{k}.
\end{align*}
Consequently,
\begin{align*}
\Tr_B S \rho^{AB} S^{\dagger} & = \sum_{j,k,\ell} c_{kjk\ell} \ket{j}\bra{\ell} = \rho^B,\\
\Tr_A S \rho^{AB} S^{\dagger} & = \sum_{i,j,k} c_{ijkj} \ket{i}\bra{k} \,\,  = \, \rho^A.
\end{align*}
Hence, $S \rho^{AB} S^{\dagger} \cv \rho^{BA} \in \Gamma^Q(\rho^B,\rho^A) = S \,\Gamma^Q(\rho^A,\rho^B) S^{\dagger}$. 
Recall that $\mathcal{H}_S,\mathcal{H}_A$ are invariant subspaces of $S$ corresponding to the eigenvalues $1$ and $-1$, respectively.  As  $C\mathcal{H}_S=0$ and $C\mathcal{H}_A\subset \mathcal{H}_A$,
it follows that $SC=CS=-C$, and thus $S C S^{\dagger} = C$. Finally, we have
\begin{align}
T^{Q}_C(\rho^A, \rho^B) & = \min_{\rho^{AB}\in \Gamma^Q(\rho^A,\rho^B)} \big( \Tr\, C \,\rho^{AB} \big) \notag \\
& = \min_{\rho^{AB}\in \Gamma^Q(\rho^A,\rho^B)} \big( \Tr\, S \,C \,S^{\dagger}\,S \, \rho^{AB} S^{\dagger} \big) \notag \\
& = \min_{\rho^{BA}\in \Gamma^Q(\rho^B,\rho^A)} \big( \Tr\, C \,\rho^{BA} \big) = T^{Q}_C(\rho^B, \rho^A). \tag*{$\Box$}
\end{align}
\end{proof}

{\bf Proposition A5}. {\sl Let $U, V \in \mathrm{U}(N)$ be unitary transformations and let $C' \vc (U \otimes V) C (U^{\dagger} \otimes V^{\dagger})$. Then, for any $\rho^A,\rho^B \in \Omega_N$,}
\begin{equation}
T^Q_C\bigl(\rho^A, \rho^B \bigr)  =  T^Q_{C'}\bigl( U\rho^AU^{\dagger}, V\rho^B V^{\dagger}\bigr).
\end{equation}
\begin{proof}\renewcommand{\qedsymbol}{}
Using the representation \eqref{rhoAB} of $\rho^{AB}$ we quickly deduce that
\begin{multline*}
(U \otimes V) \rho^{AB} (U^{\dagger} \otimes V^{\dagger}) \\
= \sum_{i,j,k,\ell} c_{ijk\ell} (U\ket{i})(V\ket{j})(\bra{k}U^{\dagger})(\bra{\ell}V^{\dagger}).
\end{multline*}
This implies that
\begin{align*}
\Tr_A \;(U \otimes V) \rho^{AB} (U^{\dagger} \otimes V^{\dagger}) = V \rho^B V^{\dagger},\\
\Tr_B \; (U \otimes V) \rho^{AB} (U^{\dagger} \otimes V^{\dagger}) = U \rho^A U^{\dagger},
\end{align*}
which proves
\begin{align*}
(U \otimes V) \, \Gamma^Q(\rho^{A},\rho^B) (U^{\dagger} \otimes V^{\dagger}) \subset \Gamma^Q(U\rho^{A}U^{\dagger},V\rho^B V^{\dagger}).
\end{align*}
Similarly, one shows 
\begin{align*}
(U^{\dagger} \otimes V^{\dagger}) \, \Gamma^Q(U\rho^{A}U^{\dagger},V\rho^B V^{\dagger}) (U \otimes V) \subset \Gamma^Q(\rho^{A},\rho^B).
\end{align*}
Hence, 
\begin{align*}
(U \otimes V) \Gamma^Q(\rho^{A},\rho^B) (U^{\dagger} \otimes V^{\dagger}) = \Gamma^Q(U\rho^{A}U^{\dagger},V\rho^B V^{\dagger}).
\end{align*}
Now, because
\begin{align*}
\Tr\, C \,\rho^{AB} = \Tr\, (U \otimes V) C (U^{\dagger} \otimes V^{\dagger}) (U \otimes V) \rho^{AB} (U^{\dagger} \otimes V^{\dagger}) 
\end{align*}
we obtain
\begin{align*}
T^{Q}_C(\rho^A, \rho^B) & = \min_{\rho^{AB}\in \Gamma^Q(\rho^B,\rho^A)} \big( \Tr\, C \,\rho^{AB} \big) \notag \\
& = \min_{\rho^{AB}\in \Gamma^Q(\rho^A,\rho^B)} \big( \Tr\, C' \, (U \otimes V) \rho^{AB} (U^{\dagger} \otimes V^{\dagger}) \big) \notag \\
& = \min_{\sigma^{AB}\in \Gamma^Q(U\rho^{A}U^{\dagger},V\rho^B V^{\dagger})} \big( \Tr\, C' \,\sigma^{AB} \big) \notag \\
&= T^{Q}_{C'}(U\rho^{A}U^{\dagger},V\rho^B V^{\dagger}). \tag*{$\Box$}
\end{align*}
\end{proof}

Formula \eqref{T_invariance} announced in the article is a simple consequence of Proposition A4.

\medskip

{\bf Corollary A6}. {\sl The optimal quantum transport cost with the cost matrix $C^Q = \tfrac{1}{2} \big(\bone_{N^2} - S\big)$ is unitarily invariant: }
\begin{equation}\label{uni}
T^Q\bigl(\rho^A, \rho^B \bigr)  =  T^Q\bigl( U\rho^AU^{\dagger}, U\rho^B U^{\dagger}\bigr),
\end{equation}
for any $U \in \mathrm U(N)$.
\begin{proof}
Observe that the cost matrix $C^Q$ is unitarily invariant, $(U \otimes U) C^Q (U^{\dagger} \otimes U^{\dagger}) = C$. Hence the result follows from Proposition A4 with $V = U$.
\end{proof}

\section{Bounds for the optimal quantum cost
\label{bounds}}

In this Section we prove the inequalities \eqref{T_bounds} in the main body of the article based on the results of \cite{YZYY18}.

Let $\rho^A$, $\rho^B$ be any two states in $\Omega_N$. Theorem 10 in \cite{YZYY18} yields
\begin{align}\label{YZYYoin}
 \tfrac{1+F(\rho^A,\rho^B)}{2} \leq \max_{\rho^{AB}\in \Gamma^Q} \Tr (\tfrac{1}{2}({\mathbbm 1}_{N^2} +S) \rho^{AB}) \leq \tfrac{1+\sqrt{F(\rho^A,\rho^B)}}{2},
\end{align}
where $S$ is the SWAP operator. Because $C^Q = \tfrac{1}{2} \big(\bone_{N^2} - S\big)$ we have
\begin{align*}
2 T^Q (\rho^A,\rho^B) =  2 \!\! \min_{\rho^{AB}\in \Gamma^Q} \big( \Tr\, C^Q \rho^{AB} \big) = 1 - \max_{\rho^{AB}\in \Gamma^Q} \big( \Tr\, S \rho^{AB} \big).
\end{align*}
Hence, from formula \eqref{YZYYoin} we deduce that
\begin{align*}
 \frac{1 - \sqrt{F(\rho^A,\rho^B)}}{2} \leq T^Q (\rho^A,\rho^B) \leq \frac{1 - F(\rho^A,\rho^B)}{2}.
\end{align*}
Furthermore, 
\begin{align*}
 1 - F(\rho^A,\rho^B) & =  \left( 1 - \sqrt{F(\rho^A,\rho^B)} \right) \left( 1 + \sqrt{F(\rho^A,\rho^B)} \right) \\
 & \leq 2 \left( 1-\sqrt{F(\rho^A,\rho^B)} \right).
\end{align*}
After taking the square root of the above inequalities we obtain the desired bounds \eqref{T_bounds}.

\medskip

We now prove yet another inequality and the last statement of Corollary 2.

\medskip

\noindent{\bf Proposition B1}. {\sl For any two states $\rho^A, \rho^B \in \Omega_N$
we have,
\begin{align}
T^Q(\rho^A,\rho^B) \leq \tfrac{1}{2} \big( 1 - \Tr \rho^A \rho^B \big).
      \label{T_ineq}
 \end{align}
Moreover, if either $\rho^A$ or $\rho^B$ is pure then}
\begin{align}
\Wq(\rho^A,\rho^B)  = \tfrac{1}{\sqrt{2}} I(\rho^A,\rho^B) = \tfrac{1}{\sqrt{2}} \sqrt{1 - \Tr \rho^A \rho^B}.
      \label{T_pure}
 \end{align}
\begin{proof}
First observe that $\rho^A \otimes \rho^B \in \Gamma^Q(\rho^A,\rho^B)$, which implies $T^Q(\rho^A,\rho^B) \leq \Tr C^Q (\rho^A \otimes \rho^B)$. Now, the well-known identity $\Tr S (\rho^A \otimes \rho^B) = \Tr \rho^A \rho^B$ (see \cite{Wer89,MPHUZ}) yields 
\begin{align*}
\Tr ({\mathbbm 1}_{N^2} - S) (\rho^A \otimes \rho^B) = 1 - \Tr \rho^A \rho^B,
\end{align*}
and formula \eqref{T_ineq} follows.

To prove the second formula we use the fact that if one of the states $\rho^A$, $\rho^B$ is pure, then $\Gamma^Q(\rho^A,\rho^B) = \{\rho^A \otimes \rho^B\}$ (as explained in \cite{FGZ19} and \cite{CEFZ21}). In such a case the inequality \eqref{T_ineq} is hence saturated.

The middle part of \eqref{T_pure} follows from the known
property of the fidelity,
 $F(\ket{\psi}\bra{\psi},\sigma) = \Tr \ket{\psi}\bra{\psi} \sigma = \bra{\psi} \sigma \ket{\psi}$.
\end{proof}

For the sake of completeness, let us also recall the Fuchs--van de Graaf inequality \cite{FvG99}, which relates the quantum fidelity to the trace distance, $D$. For any $\rho^A,\rho^B \in \Omega_N$,
\begin{align}\label{FvG}
1- \sqrt{F(\rho^A,\rho^B)} \leq D(\rho^A,\rho^B) \leq \sqrt{1- F(\rho^A,\rho^B)}.
\end{align}
This inequality, combined with Ineq.\ \eqref{T_bounds} from the main text, implies that, for any $\rho^A,\rho^B \in \Omega_N$,
\begin{align*}
W(\rho^A,\rho^B) \leq \sqrt{D(\rho^A,\rho^B)}.
\end{align*}

\section{SWAP-fidelity}\label{SWAP-f}
In this section we provide the complete proof of Proposition 1, which is based on the general results included in Section \ref{app:gen}.

For any $N$ the {\sl SWAP-fidelity}, $\fs$, is a function on $\Omega_N \times \Omega_N$ defined as follows:
\begin{equation*}
\fs(\rho^A, \rho^B) \vc
 \max_{\rho^{AB}\in \Gamma^Q(\rho^A,\rho^B)} \big( \Tr\, S \, \rho^{AB} \big).
\end{equation*}

 
\smallskip \noindent
{\bf Proposition C1}. {\sl For any two states $\rho^A, \rho^B \in \Omega_N$ we have
\begin{align*}
a) \quad & \fs(\rho^A, \rho^B) = \fs(\rho^B, \rho^A), \\
b) \quad & 0 \leq \Tr \rho^A \rho^B \leq \fs(\rho^A, \rho^B) \leq 1, \\
c) \quad & \fs(\rho^A, \rho^B) = 1 \text{ iff } \rho^A = \rho^B, \\
d) \quad & \fs(\rho^A, \rho^B) = 0 \text{ iff } \Tr \rho^A \rho^B = 0,\\
e) \quad & \fs \bigl(a \rho^A + (1-a) \sigma^A, a \rho^B + (1-a) \sigma^B \bigr) \geq \\
& \quad \geq \, a \fs \bigl(\rho^A, \rho^B \bigr) + (1-a) \fs \bigl(\sigma^A, \sigma^B \bigr), \,\text{ for } a \in [0,1], \\
f) \quad & \fs \bigl(\rho^A, \rho^B \bigr)  =  \fs \bigl( U\rho^AU^{\dagger}, U\rho^BU^{\dagger}\bigr),  \text{ for } \, U \in \mathrm U(N),\\ 
g) \quad & F(\rho^A, \rho^B) \leq \fs(\rho^A, \rho^B) \leq \sqrt{F(\rho^A, \rho^B)},\\
h) \quad & F_S \bigl(\rho^A \otimes \sigma^A, \rho^B \otimes \sigma^B \bigr) \geq F_S \bigl(\rho^A, \rho^B\bigr) F_S \bigl(\sigma^A, \sigma^B \bigr).
\end{align*}
}
\begin{proof}
By definition \eqref{cost_gen} we have, after setting $C = C^Q$, $\fs = 1 - 2 T^Q$. Consequently, points $a)$ and $c)$ follow from Theorem~A4, points $(iii)$ and $(ii)$, respectively. Similarly, point $f)$ is an immediate consequence of Corollary A6, while $e)$ follows from Proposition A3, since if $T^Q$ is convex, then $2T^Q$ is convex and  $F_S=1-2T^Q$ is concave. Point $b)$ follows from Theorem~A4 point (i) and inequality \eqref{T_ineq}. Now, point $g)$ is a straightforward implication of Ineqs.\ \eqref{YZYYoin} proven in \cite{YZYY18}. Then, point $d)$ follows from an analogous property of the quantum fidelity \cite{Jo94}.

Finally, let us address point $h)$. As previously, we set $\rho^{AB} \in \Gamma^Q(\rho^A,\rho^B)$, $\sigma^{AB} \in \Gamma^Q(\sigma^A,\sigma^B)$ to be the optimal quantum couplings
\begin{align*}
\fs(\rho^A, \rho^B) = \Tr \, S \, \rho^{AB}, && \fs(\sigma^A, \sigma^B) = \Tr \, S \, \sigma^{AB}.
\end{align*}
Now, define the following bipartite state
\begin{align*}
\tau^{AB} = U (\rho^{AB} \otimes \sigma^{AB}) U^{\dagger},
\end{align*} 
with $U: \h^A \otimes \h^B \otimes \h^A \otimes \h^B \to \h^A \otimes \h^A \otimes \h^B \otimes \h^B$ being a reshuffling operator, which acts as $U = \mathrm{id} \otimes S \otimes \mathrm{id}$. Having chosen an orthonormal basis of $\h^A \otimes \h^B$ we write 
\begin{align*}
\rho^{AB} & = \sum r_{abcd} \, \ket{a}\ket{b}\bra{c}\bra{d},\\
\sigma^{AB} & = \sum s_{e\!f\!gh} \, \ket{e}\ket{f}\bra{g}\bra{h},
\end{align*}
where the summation is over all relevant indices. Then,
\begin{align*}
\tau^{AB} & = \sum r_{abcd} \, s_{e\!f\!gh} \, \ket{a}\ket{e}\ket{b}\ket{f}\bra{c}\bra{g}\bra{d}\bra{h}.
\end{align*}
Tracing over the subsystem $A$ yields
\begin{align*}
\Tr_A \tau^{AB} & = \sum r_{abcd} \, s_{e\!f\!gh} \, \delta_{ac} \, \delta_{eg} \, \ket{b}\ket{f}\bra{d}\bra{h} \\
& = \sum r_{abad} \, s_{e\!f\!eh} \, \ket{b}\ket{f}\bra{d}\bra{h} = \rho^B \otimes \sigma^B.
\end{align*}
Analogously, one shows that $\Tr_B \tau^{AB} = \rho^A \otimes \sigma^A$, hence $\tau^{AB} \in \Gamma^{Q}(\rho^A \otimes \sigma^A,\rho^B \otimes \sigma^B)$. For such a coupling we have
\begin{align*}
\Tr S \tau^{AB} & = \Tr \, \sum r_{abcd} \, s_{e\!f\!gh} \, \ket{b}\ket{f}\ket{a}\ket{e}\bra{c}\bra{g}\bra{d}\bra{h} \\
& = \sum r_{abba} \, s_{e\!f\!f\!e} = \big( \Tr S \rho^{AB} \big) \big( \Tr S \sigma^{AB} \big) \\
& = F_S \bigl(\rho^A, \rho^B\bigr) F_S \bigl(\sigma^A, \sigma^B \bigr).
\end{align*}
Since $\Tr S \tau^{AB} \leq  F_S \bigl(\rho^A \otimes \sigma^A, \rho^B \otimes \sigma^B \bigr)$, point $h)$ follows.
\end{proof}

\section{Single-qubit transport problem: general case}
\label{single_general}

The key result in the single-qubit transport problem with cost matrix $C^Q = \tfrac{1}{2} \big({\mathbbm 1}_{4} - S\big)$ is the following:

\medskip

{\bf Theorem D1}. {\sl For any $\rho^A,\rho^B\in\Omega_2$ we have
\begin{equation}\label{mtqub1}
T^Q(\rho^A,\rho^B)= \max_{U\in\mathrm{U}(2)} \;\frac{1}{2} \left(\sqrt{(U^{\dagger}\rho^A U)_{11}}-\sqrt{(U^{\dagger}\rho^B U)_{11}}\, \right)^2,
\end{equation}
where $\rho_{11}$ denotes the upper-left entry of a $2 \times 2$ matrix $\rho$.}

\medskip

The complete proof presented in \cite{CEFZ21} relies on 
the fact that rank of an extreme point $\rho^{AB} \in \Gamma^Q(\rho^A,\rho^B)$ is always at most 2 for $\rho^A,\rho^B\in\Omega_2$
and the equivalence of the original quantum transport problem to the so-called ``dual problem'' (see also \cite{Wu19}):

\smallskip

{\bf Proposition D2}.  {\sl Let $\rho^A, \rho^B \in\Omega_N$ and let $C$ be any quantum cost matrix. Then,
\begin{multline*}
\!\!\!\!\!\! T_C^Q(\rho^A,\rho^B) =  \sup \Big\{ \Tr \big( \sigma^A  \rho^A + \sigma^B \rho^B \big) \; \Big\vert \;  \sigma^A, \sigma^B \in\mathrm{H}_N, \notag\\  C-\sigma^A\otimes \bone_N - \bone_N \otimes \sigma^B \geq 0 \Big\},
\end{multline*}
where $\mathrm{H}_N$ denotes the set of Hermitian $N \times N$ matrices.
If $\rho^A$ and $\rho^B$ are positive definite then the supremum is achieved.}

\medskip

We now introduce a convenient notation for qubits in the $y=0$ section of the Bloch ball. Let $O$ denote the rotation matrix
\begin{align*}
O(\theta)=\begin{bmatrix}\cos(\theta/2)&-\sin(\theta/2)\\\sin(\theta/2)&\cos(\theta/2)
\end{bmatrix}, \quad \text{for } \theta \in [0,2\pi),
\end{align*}
and using the Pauli matrices $\sigma_i$ define, for $r\in [0,1]$,
\begin{align*}
\rho(r,\theta) & \vc O(\theta) \begin{bmatrix}r&0\\0&1-r\end{bmatrix}O(\theta)^\top  \\
& \,= (2r-1) \big( \sin\theta \sigma_1 + \cos\theta \sigma_3 \big). 
\end{align*}

Because of unitary invariance \eqref{uni}, the quantum transport problem between two arbitrary qubits $\rho^A, \rho^B \in \Omega_2$ can be reduced to the case $\rho^A = \rho(s,0)$ and $\rho^B = \rho(r,\theta)$, with three parameters, $s,r \in [0,1]$ and $\theta \in [0,2\pi)$. The parameter $\theta$ is the angle between the Bloch vectors associated with $\rho^A$ and $\rho^B$. With such a parametrization we can further simplify the single-qubit transport problem.

\smallskip

Observe first that if $s \in \{0,1\}$ then $\rho^A$ is pure, and if $r \in \{0,1\}$ then $\rho^B$ is pure. In any such case an explicit solution of the qubit transport problem is given by formula \eqref{T_pure}. 

\medskip

{\bf Theorem D3}. {\sl Let $\rho^A = \rho(s,0), \rho^B = \rho(r,\theta)$ and assume that $0<r,s<1$. Then
\begin{align}
& T^Q(\rho^A,\rho^B) = \label{defT0}\\
&\!\! \max_{\phi\in \Phi(s,r,\theta)} \tfrac{1}{4}\big(\sqrt{1+ (2s-1)\cos\phi} - \sqrt{1+(2r-1)\cos(\theta+\phi)}\big)^2,\notag
\end{align}
where $\Phi(s,r,\theta)$ is the set of all $\phi$ satisfying the equation
\begin{equation}\label{Phieq}
\frac{(2s-1)^2\sin^2\phi}{1+(2s-1)\cos\phi}=\frac{(2r-1)^2\sin^2(\theta+\phi)}{1+(2r-1)\cos(\theta+\phi)}.
\end{equation}
Furthermore, the set $\Phi(s,r,\theta)$ has at most 6 elements.}
\begin{proof}
A unitary $2 \times 2$ matrix $U$ can be parametrized, up to a global phase, with three angles $\alpha, \beta, \phi \in [0,2\pi)$,
\begin{align*}
U = \begin{bmatrix} e^{i \alpha} & 0\\ 0 & e^{-i \alpha} \end{bmatrix} O(\phi) \begin{bmatrix} e^{i \beta} & 0\\ 0 & e^{-i \beta} \end{bmatrix}.
\end{align*}
We thus have
\begin{align*}
&\!\! (U^{\dagger}\rho(r,\theta) U)_{11} \cv f(r,\theta;\alpha,\phi) =\\
&\quad \tfrac{1}{2} \Big( 1+ (2 r-1) \big( \cos (\theta ) \cos (\phi ) + \cos (2 \alpha ) \sin (\theta ) \sin (\phi ) \big) \Big).
\end{align*}
This quantity does not depend on the parameter $\beta$, so we can set $\beta = 0$. Note also that $f(s,0;\alpha,\phi)$ does not depend on $\alpha$. With $\rho^A = \rho(s,0), \rho^B = \rho(r,\theta)$, Theorem D1 yields
\begin{align*}
T^Q(\rho^A,\rho^B) = \max_{\alpha,\phi\in [0,2\pi)} \Big( \sqrt{f(s,0;0,\phi)} - \sqrt{f(r,\theta;\alpha,\phi)} \Big)^2.
\end{align*}
Now, note that the equation $\partial_\alpha f(r,\theta;\alpha,\phi) = 0$ yields the extreme points $\alpha_0 = k \pi /2$, with $k \in \mathbb{Z}$. Since $f(r,\theta;\alpha + \pi,\phi) = f(r,\theta;\alpha,\phi)$ we can take just $\alpha_0 \in \{0,\pi/2\}$. Consequently,
\begin{align*}
T^Q(\rho^A,\rho^B) = 
\max_{\phi \in [0,2\pi)} \{ g_-(s,r,\theta;\phi), g_+(s,r,\theta;\phi) \},
\end{align*}
where we introduce the auxilliary functions
\begin{align}
& g_\pm(s,r,\theta;\phi) \vc  \label{g}\\
& \hspace*{0.9cm}\tfrac{1}{4}\Big(\sqrt{1+ (2s-1)\cos\phi} - \sqrt{1+(2r-1)\cos(\theta\pm\phi)}\Big)^2.\notag
\end{align}
But since $g_-(s,r,\theta;2\pi - \phi) = g_+(s,r,\theta;\phi)$ we can actually drop the $\pm$ index in the above formula. In conclusion, we have shown that it is sufficient to take $U = O(\phi)$ for $\phi \in [0,2\pi)$ in formula \eqref{mtqub1}.

Finally, it is straightforward to show that the equation $\partial_\phi g(s,r,\theta;\phi) = 0$ is equivalent to \eqref{Phieq}. Hence, $\Phi(s,r,\theta)$ is the set of extreme points, and \eqref{defT0} follows.

It remains to show that the set $\Phi(s,r,\theta)$ can have at most 6 elements. To this end set $z=e^{i\phi}, \zeta=e^{i\theta}$. Then \eqref{Phieq} reads
\begin{align}
&(1-2 r)^2 \left[ (2 s-1) \left(z^2+1\right)+2 z\right] \left(\zeta ^2 z^2-1\right)^2 \label{z6}\\
&\qquad -\zeta  (1-2 s)^2 \left(z^2-1\right)^2 \left[ (2 r-1) \left(\zeta ^2 z^2+1\right)+2 \zeta  z\right] = 0.\notag
\end{align}
This a 6th order polynomial equation in the variable $z$. Clearly, since we must have $\vert z \vert = 1$, not every complex root of \eqref{z6} will yield a real solution to the original  \eqref{Phieq}. Nevertheless, it can be shown  that there exist open sets in the parameter space $s,r \in (0,1)$, $\theta \in [0,2\pi)$ on which \eqref{Phieq} does have 6 distinct solutions \cite{CEFZ21}.
\end{proof}

We have thus shown that the general solution of the quantum transport problem of a single qubit with cost matrix $C^Q = \tfrac{1}{2} \big({\mathbbm 1}_{4} - S\big)$ is equivalent to solving a 6th degree polynomial equation with certain parameters. For some specific values of these parameters an explicit analytic solution can be given. This is discussed in the next two sections of this Supplemental Material. 

\section{Single-qubit transport problem: \\ 
                  commuting density matrices}
\label{single_diagonal}

A closed formula for the single-qubit optimal transport cost is available when both states are classical, i.e. represented by diagonal density matrices. 

\medskip

{\bf Proposition E1}. {\sl Let $\rho^{cl}_r={\rm \diag}(r,1-r)$ and $\rho^{cl}_s={\rm \diag}(s,1-s)$, for $r,s \in [0,1]$. Then, }
\begin{equation*}
T^Q \big(\rho^{cl}_r, \; \rho^{cl}_s\big)
= \tfrac{1}{2} \, \max\left\{ \big( \sqrt{r} - \sqrt{s} \big)^2, \big( \sqrt{1-r} - \sqrt{1-s} \big)^2 \right\}.
      \end{equation*}
\begin{proof}
Clearly, if $s=r$ then the states coincide and the transport cost vanishes. Assume then that $s\neq r$.  \eqref{Phieq} yields
\begin{equation}\label{Phieq_diag}
\frac{(2s-1)^2\sin^2\phi}{1+(2s-1)\cos\phi}=\frac{(2r-1)^2\sin^2 \phi }{1+(2r-1)\cos \phi}.
\end{equation}
This shows that $0, \pi \in \Phi(r,s,0)$, and both are double roots of  \eqref{Phieq_diag}. Then  \eqref{g} then immediately yields
\begin{align*}
g(r,s,0;0) & = \tfrac{1}{2} \,\big( \sqrt{r} - \sqrt{s} \big)^2, &&\\
g(r,s,0;\pi) & = \tfrac{1}{2} \,\big( \sqrt{1-r} - \sqrt{1-s} \big)^2.&&
\end{align*}

There is yet another solution to  \eqref{Phieq_diag}, which reads
\begin{align}\label{phi0}
\cos\phi_0 = \frac{2(1-r-s)}{(2r-1)(2s-1)}.
\end{align}
Note, however, that the absolute value of the RHS of  \eqref{phi0} can become larger than 1 when $s$ or $r$ is close to 1/2. In either case, one can quickly convince oneself 
that if $\phi_0 \in [0,2\pi)$, then actually $g(r,s,0;\phi_0) \leq g(r,s,0;0)$ and $g(r,s,0;\phi_0) \leq g(r,s,0;\pi)$.
\end{proof}

An alternative proof based on the classical transport problem is provided
in \cite{CEFZ21}.

\medskip

Assume now that one of the states is maximally mixed,  say $\rho^A = \tfrac{1}{2} \bone_2$, and let $U \in \mathrm U(2)$ be such that $U \rho^B U^{\dagger} = \diag(\lambda,1-\lambda)$. Then
\begin{align*}
T^Q(\rho^A,\rho^B) &= T^Q(\rho^A,U \rho^B U^{\dagger}) = T^Q(\rho^{cl}_{1/2},\rho^{cl}_{\lambda}) \\
& = \tfrac{1}{4} \max \big\{ \big(1-\sqrt{2 \lambda} \big)^2, \big(1-\sqrt{2 (1-\lambda)}\big)^2 \big\}.
\end{align*}
This fact implies Eq.\ \eqref{T_mixed} in the main body of the article.

\medskip

Formula \eqref{WB} presented in the article is a simple consequence of Proposition E1. For a vector $\vec{\tau} \in \mathbb{R}^3$ with $\Vert \vec{\tau} \Vert \in [0,1]$ define $\rho_\pm(\vec{\tau}) \vc \tfrac{1}{2} \left( {\mathbbm 1} \pm \vec{\tau} \cdot \vec{\sigma} \right) \in \Omega_2$, with the Pauli matrices $\sigma_i$. 

\smallskip
{\bf Corollary E2}. We have 
\begin{align*}
\Wq \bigl(\rho_+(\vec{\tau}),\rho_-(\vec{\tau}\bigr) = 
\tfrac{1}{\sqrt{2}} B \bigl(\rho_+(\vec{\tau}),\rho_-(\vec{\tau})\bigr) = \sqrt{1-\sqrt{1-\Vert \vec{\tau} \Vert^2}}.
\end{align*}
\begin{proof} For  any $\vec{\tau}$ the density matrices $\rho_\pm(\vec{\tau})$ commute and hence can be simultaneously diagonalized. Let us denote by $r \vc \tfrac{1}{2} \big( 1 + \Vert \vec{\tau} \Vert \big)$. Then, by unitary invariance of $T^Q$, we have 
\begin{align*}
T^Q \bigl(\rho_+(\vec{\tau}),\rho_-(\vec{\tau}\bigr) = T^Q \bigl(\rho^{cl}_r,\rho^{cl}_{1-r}\bigr).
\end{align*}
Proposition E1 yields
\begin{align*}
T^Q \bigl(\rho^{cl}_r,\rho^{cl}_{1-r}\bigr) = 1-2 \sqrt{r(1-r)} = 1-\sqrt{1-\Vert \vec{\tau} \Vert^2}.
\end{align*}
On the other hand, exploiting the fact that quantum fidelity is also unitary invariant, we obtain
\begin{align*}
F \bigl(\rho_+(\vec{\tau}),\rho_-(\vec{\tau}\bigr) = F \bigl(\rho^{cl}_r,\rho^{cl}_{1-r}\bigr) = \sqrt{1-\Vert \vec{\tau} \Vert^2}.
\end{align*}
Since $B(\rho^A,\rho^B) = \sqrt{2(1-\sqrt{F(\rho^A,\rho^B)}}$, the assertion follows.
\end{proof}

\section{Single-qubit transport problem: \\
              two isospectral density matrices}
\label{single_isospectral}

In this section we prove, using Theorem D3, formula \eqref{T_isosp} announced in the main text. Assume that the spectra of $\rho^A, \rho^B \in \Omega_2$ are equal. Because of unitary invariance \eqref{uni} and using the parametrization introduced in Sec.~\ref{single_general}, without loss of generality we can set $\rho^A = \rho(s,0)$ and $\rho^B = \rho(s,\theta)$ for some $s \in [0,1]$ and $\theta \in [0,2\pi)$. Then the following result implies \eqref{T_isosp}.

\medskip

{\bf Theorem F1}. {\sl For any $s \in [0,1]$ and $\theta \in [0,2\pi)$ we have}
\begin{align}\label{Tiso}
T^Q \big(\rho(s,0),\rho(s,\theta) \big) = \Big( \tfrac{1}{2} -\sqrt{s(1-s)} \Big) \sin^2 (\theta/2).
\end{align}

\begin{proof}
Note first that if the states $\rho^A,\rho^B$ are pure, i.e. $s = 0$ or $s=1$, formula \eqref{Tiso} gives $T^Q \big(\rho(s,0),\rho(s,\theta) \big) = \tfrac{1}{2} \sin^2 (\theta / 2)$, which agrees with  \eqref{T_pure}.

From now on we assume that that $\rho^A, \rho^B$ are not pure.  When $r = s$, \eqref{z6} simplifies to the following:
\begin{align}
& (\zeta -1) (1-2 s)^2 \left(\zeta  z^2-1\right) \times \label{zeta_iso} \\
&\quad \times  \left[4 s (\zeta +1) \left(\zeta  z^2+1\right) z +(2 s-1) (z-1)^2 (\zeta  z-1)^2 \right]= 0.\notag
\end{align}

Eq.\ \eqref{zeta_iso} is satisfied when $z = \pm \zeta^{-1/2}$. This corresponds to $\phi_0 = -\theta/2$ or $\phi_0' = \pi - \theta/2$. Observe, however, that we have $g(s,s,\theta;\phi_0) = g(s,s,\theta;\phi_0') = 0$, so we can safely ignore $\phi_0, \phi_0' \in \Phi(s,s,\theta)$ in the maximum in  \eqref{defT0}.

Hence, we are left with a 4th order equation
\begin{align}\label{zeta4}
4 s (\zeta +1) \left(\zeta  z^2+1\right) z +(2 s-1) (z-1)^2 (\zeta  z-1)^2 = 0,
\end{align}
which, converting back to the variables $\theta,\phi$, reads
\begin{multline}\label{phi4}
(2 s-1) \big[ 2 + \cos (\theta +2 \phi )+ \cos (\theta ) \big] +  \\
\qquad +2 \big[ \cos (\theta +\phi )+ \cos (\phi ) \big] = 0.
\end{multline}
Now, observe that if $\phi$ satisfies  \eqref{phi4}, then so does $\phi' = -\phi - \theta$. This translates to the fact that if $z$ satisfies  \eqref{zeta4}, then so does $(z \zeta)^{-1}$. Furthermore, $g(s,s,\theta;\phi) = g(s,s,\theta;\phi')$. Hence, in the isospectral case we are effectively taking the maximum over just two values of $\phi$.

Let us now seek an angle $\phi_1 \in [0,2\pi)$ such that $g(s,s,\theta;\phi_1)$ equals the RHS of  \eqref{Tiso}. The latter equation reads
\begin{align*}
& \Big\{ (2 s-1) \big[\cos \left(\theta +\phi _1\right)+\cos \left(\phi _1\right)\big] \\
& \quad -\big(2 \sqrt{s(1-s)}-1\big) \big(\cos (\theta )-1\big)+2\Big\}^2 \\
& \qquad\quad = 4 \big[(2 s-1) \cos \left(\phi _1\right)+1\big] \big[(2 s-1) \cos
   \left(\theta +\phi _1\right)+1\big].
\end{align*}
In terms of $z$ and $\zeta$, the above is equivalent to a 4th order polynomial equation in $z$, which can be recast in the following form:
\begin{align}\label{zeq}
\Big[ \zeta  (1-2 s) z^2+(\zeta +1) \big(2 \sqrt{s(1-s)}-1\big) z-2 s+1 \big]^2 = 0.
\end{align}
Hence,  \eqref{zeq} has two double roots:
\begin{align*}
& z_1^{\pm} = \big[ 2 \zeta  (1-2 s) \big]^{-1} \bigg\{ (\zeta +1) \big( 1-2 \sqrt{s(1-s)} \, \big) \\
& \hspace*{1.5cm} \pm \sqrt{(\zeta +1)^2 \big(1-2 \sqrt{s(1-s)} \,\big)^2-4 \zeta  (1-2s)^2} \bigg\}.
\end{align*}
Furthermore, one can check that $z_1^{-} = (\zeta z_1^{+})^{-1}$. Now, it turns out that $z_1^{\pm}$ are also solutions to \eqref{zeta4}.
We thus conclude that $\phi_1, \phi_1' \in \Phi(s,s,\theta)$.

We now divide the polynomial in \eqref{zeta4} by $(z-z_1^{+})(z-z_1^{-})$. We are left with the following quadratic equation
\begin{align*}
\zeta  \Big[ (2 s-1) \left(\zeta  z^2+1\right)+(\zeta +1) \big(2 \sqrt{(1-s) s}+1\big) z\Big] = 0.
\end{align*}
Its solutions are
\begin{align*}
& z_2^{\pm} = \big[ 2 \zeta  (1-2 s) \big]^{-1} \bigg\{ (\zeta +1) \big( 1+2 \sqrt{s(1-s)} \, \big) \\
& \hspace*{1.5cm} \pm \sqrt{(\zeta +1)^2 \big(1+2 \sqrt{s(1-s)} \,\big)^2-4 \zeta  (1-2s)^2} \bigg\}.
\end{align*}
Again, we have $z_2^{-} = (\zeta z_2^{+})^{-1}$, in agreement with the symmetry argument. Setting $z_2^+ \cv e^{i \phi_2}$ and $z_2^- \cv e^{i \phi_2'}$ we have $\phi_2, \phi_2' \in \Phi(s,s,\theta)$. Then we deduce
\begin{align*}
g(s,s,\theta;\phi_2) & = g(s,s,\theta;\phi_2') \\
& \hspace*{-0.4cm} = \tfrac{1}{4} \Big[ (1-6 \sqrt{(1-s) s} - \big(1+2 \sqrt{(1-s) s} \, \big) \cos (\theta ) \Big].
\end{align*}

Finally, we observe that
\begin{align*}
g(s,s,\theta;\phi_1) - g(s,s,\theta;\phi_2) = \sqrt{(1-s) s} \, \big(1+\cos (\theta ) \big) \geq 0.
\end{align*}
This shows that, for any $s \in (0,1)$, $\theta \in [0,2\pi)$,
\begin{align*}
T^Q \big(\rho(s,0),\rho(s,\theta) \big) = g(s,s,\theta;\phi_1),
\end{align*}
and \eqref{Tiso} follows.
\end{proof}

Note that $g(s,s,\theta;\phi_2)$ can become negative for certain values of $s$ and $\theta$. This means that for such values $\Phi(s,s,\theta) = \{\phi_0,\phi_0',\phi_1,\phi_1'\}$.

\section{The triangle inequality for the transport distance}
\label{triangle}

Given Theorem D1, the triangle inequality for $\Wq = \sqrt{T^Q}$ comes as an immediate corollary.

Indeed, let $\rho^A, \rho^B, \rho^C \in \Omega_2$, and let $U_0 \in \mathrm U(2)$ denote the unitary matrix which gives the maximum of $T^Q(\rho^A,\rho^C)$ in  \eqref{mtqub1}. Then we have
\begin{align*}
\Wq(\rho^A,\rho^C) & = \tfrac{1}{\sqrt{2}} \Big\vert \sqrt{(U_0^{\dagger}\rho^A U_0)_{11}}-\sqrt{(U^{\dagger}_0\rho^B U_0)_{11}}\, \Big\vert \\
& \leq \tfrac{1}{\sqrt{2}} \Big\vert \sqrt{(U_0^{\dagger}\rho^A U_0)_{11}}-\sqrt{(U^{\dagger}_0\rho^B U_0)_{11}}\, \Big\vert \\
& \qquad + \tfrac{1}{\sqrt{2}} \Big\vert \sqrt{(U_0^{\dagger}\rho^B U_0)_{11}}-\sqrt{(U^{\dagger}_0\rho^C U_0)_{11}}\, \Big\vert \\
& \leq \Wq(\rho^A,\rho^B) + \Wq(\rho^B,\rho^C).
\end{align*}

Recall also that if $W$ is a metric, then so is $h(W)$ for any concave function $h$. Since $\Wq_p = (T^Q)^{1/p} = \Wq^{2/p}$ and $x^{2/p}$ is a concave function on $\mathbb{R}^+$ for $p \geq 2$, we conclude that $\Wq_p$ is a distance on $\Omega_2$ for any $p \geq 2$. 

On the other hand, we stress that $W_1 = T^Q$ does \emph{not} satisfy the triangle inequality,
and thus it is not a distance on qubits, but only a semidistance. Furthermore, 
it is possible to show \cite{CEFZ21} that the triangle inequality also fails for $p \in (1,2)$.

\section{Quantum-to-classical transition of the optimal transport problem}
\label{app:decoh}

In this section we study the quantum-to-classical transition of the transport problem. We assume that decoherence occurs through a dephasing channel (cf. \cite{Zurek03,Pre18}). The latter is a superoperator $\mathcal{E}_{\alpha}$, parametrized by $\alpha \in [0,1]$, which acts as
\begin{align*}
\mathcal{E}_{\alpha}(A) = \alpha A + (1-\alpha) \diag (A),
\end{align*}
for any matrix $A \in \mathcal{B}(\mathbb{C}^N) \simeq \mathbb{C}^{N \times N}$. For $\alpha=1$, the channel $\mathcal{E}_1$ is the identity, while for $\alpha=0$ it gives a diagonal matrix. For a state $\rho \in \Omega_N$, the parameter $\alpha$ is proportional to the $l_1$-coherence \cite{BCP14} of the state $\rho_{\alpha} \vc \mathcal{E}_{\alpha}(\rho)$. In terms of physical models, the decoherence parameter is time dependent: $\alpha = e^{-\Gamma t}$, where $\Gamma$ characterizes the interaction of the system with an environment, e.g. through scattering processes.

Suppose first that the input states $\rho^A, \rho^B \in \Omega_N$ suffer from decoherence, while the cost matrix $C^Q$ is fixed. Then,

\smallskip

{\bf Proposition H1} {\sl The optimal quantum transport cost between two density matrices $\rho_{\alpha}^A \neq \rho_{\alpha}^B \in \Omega_N$ decreases with the parameter $\alpha$,}
\begin{align*}
T^Q (\rho_{\alpha}^A, \rho_{\alpha}^B) \leq T^Q (\rho_{\beta}^A, \rho_{\beta}^B), && \text{for} && 0\leq \alpha \leq \beta \leq 1.
\end{align*}
\begin{proof}
We first show that
\begin{align}\label{coh1}
T^Q \big( \diag(\rho^A),\diag(\rho^B) \big) \leq T^Q(\rho^{A},\rho^B).
\end{align}

To this end let $\mathcal{D}_N \subset \mathrm{U}(N)$ be the subgroup of diagonal matrices with diagonal entries $\pm 1$. Proposition A5 yields $T^Q(\rho^{A},\rho^B) = T^Q(D \rho^{A}D^{\dagger},D\rho^B D^{\dagger})$ for any $D \in \mathcal{D}_N$. Recall also that $|\mathcal{D}_N|=2^N$. Now note that
\begin{equation*}
\diag(\rho)=2^{-N}\sum_{D\in\mathcal{D}_N} D\rho D^{\dagger},
\end{equation*}
for any $\rho \in \Omega_N$. Then the convexity of $T^Q$ (recall Proposition A3) implies
\begin{align*}
T^Q \big( \diag(\rho^A),\diag(\rho^B) \big) & \leq 2^{-N}\sum_{D\in\mathcal{D}_N} T^Q(D\rho^{A}D^{\dagger},D\rho^BD^{\dagger}) \\
& = T^Q(\rho^{A},\rho^B).
\end{align*}

Now assume that $0 \leq \alpha \leq \beta \leq 1$. We can then write
\begin{align*}
\rho_{\alpha} & \vc \alpha \rho + (1-\alpha) \diag(\rho) \\
& = \alpha \big[ \beta^{-1} \rho_\beta + (1- \beta^{-1}) \diag (\rho) \big] + (1-\alpha) \diag(\rho) \\
& = \alpha \beta^{-1} \rho_{\beta} + (1- \alpha \beta^{-1}) \diag (\rho).
\end{align*}
Because $\alpha \beta^{-1} \in [0,1]$ we can again invoke the convexity of $T^Q$ to conclude that
\begin{align*}
T^Q(\rho^{A}_\alpha,\rho^B_\alpha) &\leq \alpha \beta^{-1} T^Q(\rho^{A}_\beta,\rho^B_\beta) \\
& \qquad + (1- \alpha \beta^{-1}) T^Q \big( \diag(\rho^A),\diag(\rho^B) \big) \\
& \leq T^Q(\rho^{A}_\beta,\rho^B_\beta).
\end{align*}
The last inequality follows from \eqref{coh1} because $\diag(\rho_\beta) = \diag(\rho)$.
\end{proof}

\medskip

Suppose now that decoherence affects the quantum cost matrix $C$. Note, however, that $C_{\alpha} = \mathcal{E}_\alpha(C)$ is \emph{not} a quantum cost matrix (recall Def.\ A1) for $\alpha < 1$. On the other hand, $C_0$ is a diagonal matrix which can be identified with the cost matrix of the corresponding classical problem. The transition between the quantum and classical optimal transport problem can be studied with the help of the function $T^Q_{C,\alpha} \vc \min_{\rho^{AB}\in \Gamma^Q} \big( \Tr\, \mathcal{E}_{\alpha}(C) \rho^{AB} \big)$.

We now focus on the case $N=2$, $C = C^Q = \tfrac{1}{2} \big(\bone_4 -~S\big)$ and denote $T^Q_{\alpha} \vc T^Q_{C^Q,\alpha}$ For more general results, the reader is invited to consult \cite{CEFZ21}. 

Let us denote by $\Omega_2^{cl}$ the subset of all commuting density matrices of order 2. We have the following result:

\medskip

{\bf Theorem H2}. {\sl For any $r,s \in [0,1]$ let $s_0,r_0$ be defined by
\begin{align*}
(\sqrt{r_0} - \sqrt{s_0})^2 \vc \max\left\{ \big( \sqrt{r} - \sqrt{s} \big)^2, \big( \sqrt{1-r} - \sqrt{1-s} \big)^2 \right\}.
\end{align*}
Then for any $\rho^{cl}_r,\rho^{cl}_s \in \Omega_2^{cl}$,}
\begin{align}
& T^Q_{\alpha}(\rho^{cl}_r,\rho^{cl}_s)= \label{Talpha}\\
& \quad\; \begin{cases}
\tfrac{1}{2} \sqrt{1-\alpha^2} \, \vert r_0-s_0 \vert,  & \text{ for } 0\le \alpha<\tfrac{2\sqrt{r_0 s_0}}{r_0 + s_0},\\
\tfrac{1}{2} (\sqrt{r_0} - \sqrt{s_0})^2+(1-\alpha)\sqrt{r_0 s_0}, & \text{ for } \tfrac{2\sqrt{r_0 s_0}}{r_0 + s_0}\le \alpha\le 1.
\end{cases} \notag
\end{align}

\smallskip

This result can be derived as a variant of the classical transport problem \cite{CEFZ21}.

\medskip

From \eqref{Talpha} it is obvious that $T^Q_{\alpha}$  is positive, symmetric and vanishes if and only if $r=s$. One can also quickly check that $\Wq^{\alpha} = \sqrt{T^Q_{\alpha}}$ obeys the triangle inequality. Hence, $\Wq^{\alpha}$ is actually a distance on $\Omega_2^{cl}$ for any $\alpha \in [0,1]$.

Observe also that if either of the states, say $\rho^{cl}_r$, is pure, then $r_0 = 0$ and the transport cost $T^Q_{\alpha}(\rho^{cl}_r,\rho^{cl}_s)$ does not depend on $\alpha$. In particular, for $\rho^{cl}_r$ pure we have
\begin{align*}
T^Q(\rho^{cl}_r,\rho^{cl}_s) = T^{cl}(\rho^{cl}_r,\rho^{cl}_s).
\end{align*}

Formula \eqref{Talpha} also implies that if neither of the classical states is pure and $r \neq s$, then $T^Q_{\alpha}$ is a strictly decreasing function of $\alpha$:
\begin{align*}
\alpha < \beta \quad \Rightarrow \quad T^Q_{\alpha}(\rho^{cl}_r,\rho^{cl}_s) > T^Q_{\beta}(\rho^{cl}_r,\rho^{cl}_s).
\end{align*}
This shows that decoherence always increases the cost of the optimal transport. 

\begin{figure}[h]
	\centering
	\includegraphics[width=1\linewidth]{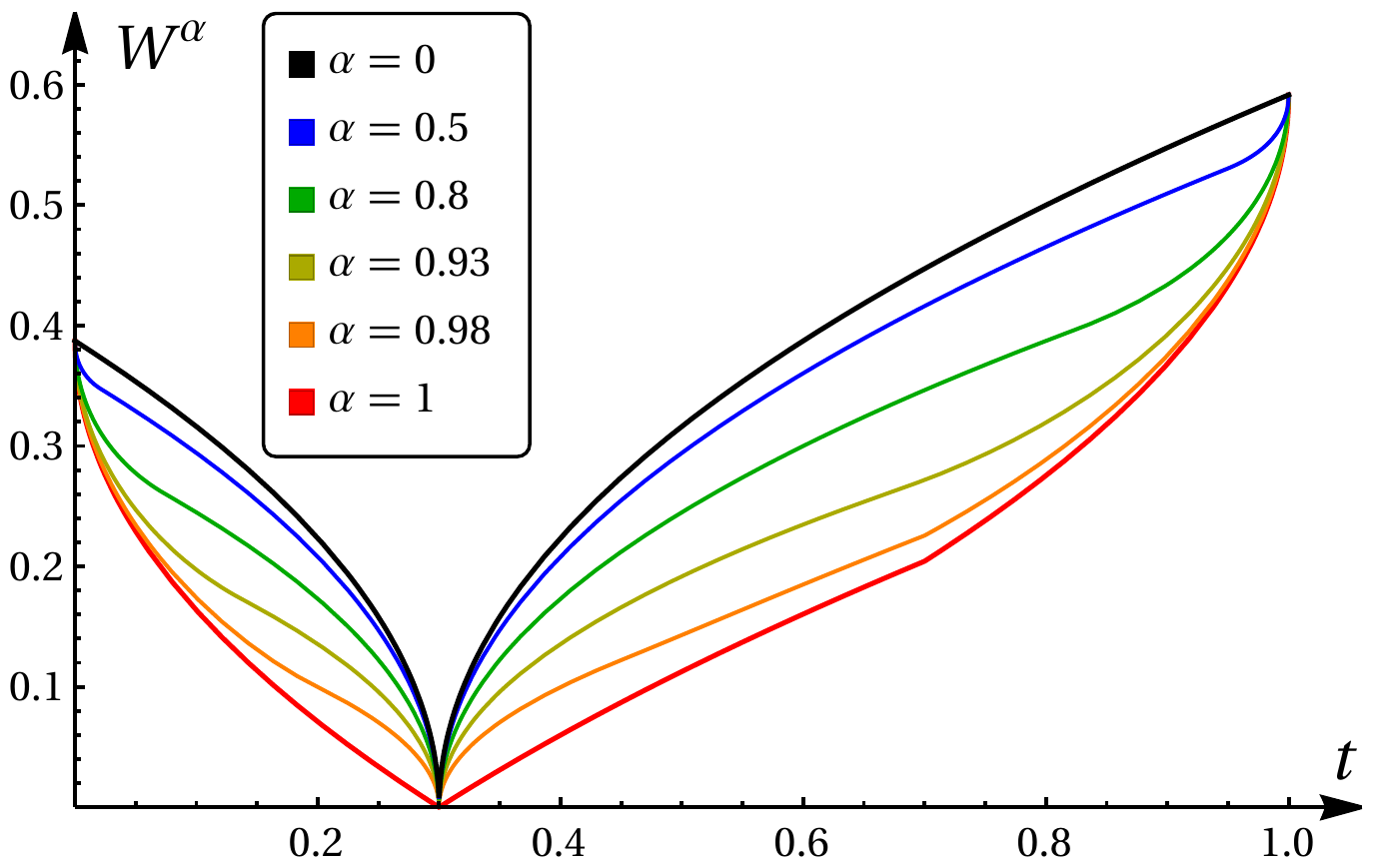}
	 \caption{An illustration of quantum-to-classical transition of the transport problem. The plot shows
	 how the distance 
$\Wq^{\alpha}(\diag(0.3,0.7),\diag(t,1-t))$ increases with the coherence parameter
decreasing from $\alpha=1$ (quantum) to $\alpha=0$ (classical) case.}
        \label{fig4}
\end{figure}

For $\alpha = 0$ we have $\Wq^0(\diag(r,1-r),\diag(s,1-s)) = \sqrt{\vert r - s\vert /2}$, which is the classical 2-Wasserstein distance \cite{Maas2011} between probability vectors $p^A = (r,1-r)$ and $p^B=(s,1-s)$.  Hence, $W^{\alpha}$ interpolates continuously between $\Wq^0 = \Wq^{cl}$ and its quantum analogue $\Wq^1 = \Wq$. 
Furthermore, for two classical mixed states $\rho^A \neq \rho^B$, 
the distance
$\Wq^{\alpha}(\rho^A,\rho^B)$ is a strictly decreasing function of $\alpha$. The full pattern of decoherence of the quantum optimal transport, illustrated in Fig.\ \ref{fig4}, is rather involved.

\section{\label{app:cost_E}Cost matrices for larger dimensions}
\label{cost_dim}

Let us discuss here the transport distance between  two 
mixed states of size $N> 2$. In the case of a qutrit ($N=3$)
the quantum cost matrix (Eq.\ \eqref{costLn} in the main body of the article) for the configuration of the  equilateral triangle, 
$E_{12}=E_{13}=E_{23}=1$,
takes the form
\begin{equation}
C^Q   
 \vc \frac{1}{2}
  \left[
\begin{array}{lllllllll}
   0  &  0 & 0 & 0 & 0 & 0 & 0 & 0 &0  \\
   0  &  1 & 0 & \!\!\!\!-1 & 0 & 0 & 0 & 0 &0  \\
   0  &  0 & 1 & 0 & 0 & 0 &  \!\!\!\!-1 & 0 &0  \\
   0  &   \!\!\!\!-1 & 0 & 1 & 0 & 0 & 0 & 0 &0  \\
   0  &  0 & 0 & 0 & 0 & 0 & 0 & 0 &0  \\
   0  &  0 & 0 & 0 & 0 & 1 & 0 &  \!\!\!\!-1 &0  \\
   0  &  0 &  \!\!\!\!-1 & 0 & 0 & 0 & 1 & 0 &0  \\
   0  &  0 & 0 & 0 & 0 &  \!\!\!\!-1 & 0 & 1 &0  \\
   0  &  0 & 0 & 0 & 0 & 0 & 0 & 0 &0  
\end{array}
\right] .
\label{cost3}
\end{equation}

Observe that the above matrix can be taken to
block-diagonal form  by a suitable permutation.
Note that the corresponding  classical cost function,
$C^{cl}= {\rm diag}(C^Q)$, in two-index notation, reads
$C^{cl}_{ij,ij}=1$ if $i \ne j$ 
and $C^{cl}_{ii,ii}=0$. 

Numerical results suggest that, as in the qubit case, the root
transport cost for qutrits, $\Wq = \sqrt{T^Q}$,
 satisfies the triangle inequality, while the latter fails for $\Wq_1 = T^Q$. We have also checked that this extends to ququarts ($N = 4$), 
 with $C^Q = \tfrac{1}{2} \big( \bone_{16} - S \big)$.
 
It is also natural to consider a set of three
ordered points on a line equipped with the Euclidean distance, $E_{12} = E_{23} =~1, E_{13} = 2$. The corresponding classical cost matrix is defined as
      \begin{equation}
      (C^{cl}_E)_{ij,ij}=|i-j|, \quad i,j=1,\dots, N.
      \label{costL2}
      \end{equation}
The quantization (Eq.\ \eqref{costWnbis} in the main body of the article) of such $C^{cl}_E$ reads
\begin{equation}
C^{Q}_E   \vc \frac{1}{2}
  \left[
\begin{array}{lllllllll}
   0  &  0 & 0 & 0 & 0 & 0 & 0 & 0 &0  \\
   0  &  1 & 0 & \!\!\!\!-1 & 0 & 0 & 0 & 0 &0  \\
   0  &  0 & 2 & 0 & 0 & 0 &  \!\!\!\!-2 & 0 &0  \\
   0  &   \!\!\!\!-1 & 0 & 1 & 0 & 0 & 0 & 0 &0  \\
   0  &  0 & 0 & 0 & 0 & 0 & 0 & 0 &0  \\
   0  &  0 & 0 & 0 & 0 & 1 & 0 &  \!\!\!\!-1 &0  \\
   0  &  0 &  \!\!\!\!-2 & 0 & 0 & 0 & 2 & 0 &0  \\
   0  &  0 & 0 & 0 & 0 &  \!\!\!\!-1 & 0 & 1 &0  \\
   0  &  0 & 0 & 0 & 0 & 0 & 0 & 0 &0  
\end{array}
\right] ,
\label{cost3b}
\end{equation}
so that 
$C^{Q}_E\ge 0$ and  diag$(C^{Q}_{E})=C^{cl}_E$.

Under such a choice the quantum Kantorovich--Wasserstein distance is no longer unitarily invariant, 
 as the distance from $|1\rangle$ to $|3 \rangle$
is larger than from  $|1\rangle$ to $|2 \rangle$,
but such a property is desirable to get the
 correct classical limit \cite{ZS98,ZS01}.

To generalize this expression for an arbitrary dimension $N$
we can use the notion of maximally entangled states   
$|\psi^-_{ij}\rangle$ which act on a two dimensional subspace. 
The corresponding cost matrix for the transport problem 
is then given by a weighted combination of projections onto antisymmetric subspaces,
\begin{equation}
 C^Q_E =  
 \sum_{j>i=1}^{N}  \vert i-j \vert \;  |\psi^-_{ij}\rangle  \langle \psi^-_{ij}|.
      \label{costWn}
      \end{equation}
      
Observe that, in contrast to the simplex case, such $C^Q_E$ is no longer a projection, and the quantum transport cost $T_{E,p}^Q$ does depend on the choice of parameter $p \geq 1$. We have checked numerically that while $\sqrt{T_{E,1}^Q}$ seems to enjoy the triangle inequality, this is not the case for either $\sqrt{T_{E,2}^Q}$ or $T_{E,1}^Q$.

Based on numerical results we are tempted  to conjecture that for any classical geometry $E$ on the $N$-point space and any three quantum states 
$\rho^A, \rho^B, \rho^C \in \Omega_N$
one has
\begin{align*}
\sqrt{T^Q_{E,1}(\rho^A, \rho^B)}+\sqrt{T^Q_{E,1}(\rho^A, \rho^B)}\ge \sqrt{T^Q_{E,1}(\rho^A, \rho^B)}.
\end{align*}

\medskip

Moreover, the two quantities
$\sqrt{T_{E,2}^Q}$ and $T_{E,1}^Q$
 give distances between basis states $|i\rangle$ and $|j\rangle$
 consistent with the classical distance matrix $E_{ij}$.

\section{Coupling matrices in the Fano form}
\label{App_Fano}
When analyzing density matrices of order $N=2$
it is convenient to use the set of
three Pauli matrices, 
   which generate the group $\mathrm{SU}(2)$.
   These traceless matrices, often denoted as $\sigma_1, \sigma_2 $ and $\sigma_3$,
  together with the identity matrix, $\sigma_0:={\mathbbm 1}_2$, form an 
   orthogonal basis
   in the Hilbert-Schmidt  space of Hermitian matrices of dimension $2$. Hence, any $2 \times 2$ Hermitian matrix $\rho$
       can be expanded in this basis,
       \begin{equation}
     \rho= \frac{1}{2}  \sigma_0 + \frac{1}{2} \sum_{i=1}^3 \tau_i \sigma_i,
    \label{Bloch}
      \end{equation}
     where the  expansion coefficients are given by  $\tau_i = {\rm Tr} \rho \sigma_i$.
     Since the state $\rho$ is Hermitian  these three numbers are real.
     The vector $\tau$ of length three is called  the {\sl Bloch vector} of $\rho$,
       and if its length  satisfies condition $\Vert \tau \Vert^2\le 1$
, then
         the matrix $\rho$ is positive semidefinite and represents a legitimate quantum state \cite{BZ17}.
     
     In the general case of a state $\rho$ of dimension $N$ the generalized Bloch vector $\tau$
     consists of $N^2-1$ components, and the set of three Pauli matrices is
      replaced by the collection of $N^2-1$ traceless Hermitian matrices $\Lambda_i$
      which satisfy the orthogonality relation Tr$\Lambda_i \Lambda_j = 2 \delta_{ij}$
        and   generate the group  $\mathrm{SU}(N)$.  This yields the expansion
  \begin{equation}
\rho= \frac{1}{N}  {\mathbbm 1}_N+ \frac{1}{N} \sum_{i=1}^{N^2-1} \tau_i \Lambda_i. 
   \label{Bloch_n}
      \end{equation}

 Usually the order of the generators is not relevant, 
 but for our purposes it is convenient to select first  $N-1$
generators $\Lambda_i$ as diagonal ones.
Then any classical probability vector $p$ of size $N$ from the probability simplex, 
$\Delta_N$, 
can be expressed in its {\sl Bloch form},
which can be considered as a special case of  formula  (\ref{Bloch_n}) above,
   \begin{equation}
     p= \frac{1}{N}\;  {\rm diag}(  {\mathbbm 1}_N) + \frac{1}{N} \sum_{i=1}^{N-1} \tau_i \; {\rm diag}(\Lambda_i).
    \label{Bloch_class}
      \end{equation}
Here the first $N - 1$ generators $\{\Lambda_i\}_{i=1}^{N-1}$
  are represented
 by diagonal traceless matrices of dimension $N$.
 The first term simply represents the flat vector, $p_*=(1,\dots,1)/N$,
 while the second one describes the translation vector $\tilde \tau$
 inside the simplex, which consists of the first $N-1$ components 
 of the Bloch vector $\tau$ of length $N^2-1$.
      
\medskip

   Now consider an arbitrary state  $\rho^{AB}$ 
  of a bipartite $N \times N$ system.
   In full analogy with the Bloch representation (\ref{Bloch_n}), 
      one  can expand it in the product basis, $\Lambda_i \otimes \Lambda_j$
   with $i,j=0,\dots N^2-1$. In this way one arrives at the 
   {\sl Fano representation} \cite{Fa83} of a
   bipartite state,
 \begin{equation}
   \rho^{AB}= \frac{1}{N^2}  \sum_{i,j=0}^{N^2-1} M_{ij} \Lambda_i \otimes \Lambda_j,
    \label{Fano_n}
      \end{equation}
    where $M_{ij}= \frac{N^2}{4} {\rm Tr} \rho^{AB} \Lambda_i \otimes \Lambda_j$.
  Since we have selected 
   $\Lambda_0 ={\mathbbm 1}_{N^2}$,
   the matrix of coefficients $M_{ij}$ takes the form
   \begin{equation}
 M  =
  \left[
\begin{array}{ll}
   1  & \ a^T  \\
  b  & \   R  \\ 
\end{array}
\right] , \ 
\label{Fano2}
\end{equation}
where $R$ is a real correlation matrix of order $N^2-1$,
while the vectors $a$ and $b$ of length $N^2-1$ determine the Bloch vectors
of both partial traces,
$a=\tau_A$,  $\rho^A={\rm Tr}_B \rho^{AB}$,
and
$b=\tau_B$,
$\rho^B={\rm Tr}_A \rho^{AB}$,
respectively.
This representation is useful to determine the
maximal fidelity of a given two-qubit state with respect to maximally
entangled states \cite{BH300},
and to formulate separability criteria for bipartite systems \cite{Vi07}. 
The matrix (\ref{Fano2}) allows us to represent any bipartite state
by specifying both local states and their correlations,
$\rho^{AB}= \rho^{AB}(\tau_A, \tau_B, R)$.

In the special  case of a diagonal state $\rho^{AB}$,
any bipartite 
probability vector $p^{AB} \in \Delta_{N^2}$ 
can be written as 
$p^{AB}= p^{AB}(\tilde{\tau_A}, \tilde{\tau_B}, \tilde{R})$.
Here $\tilde{\tau_A}$ and  $\tilde{\tau_B}$
denote vectors formed from the first $N-1$ components of the Bloch vectors
 $\tau_A$ and  $\tau_B$, respectively,
 which, according  to  (\ref{Bloch_class}),
  determine both marginal probability 
 vectors, $p^A$ and $p^B$.
 The classical correlation matrix  $\tilde R$ of order $N-1$
forms the upper left corner of the  
full correlation matrix $R$ of order $N^2-1$ in  (\ref{Fano2}). 

\medskip

 In the one-qubit case, $N=2$, the real  matrix $R$ of order three can be 
 taken to diagonal form via real singular value
 decomposition, $R \to R'=O_1RO_2$ where $O_1,O_2 \in SO(3)$,
 so that $R'$ is diagonal and its entries $v_1,v_2,v_3$, are real and can be negative. 
 If both partial traces of $\rho^{AB}$ are maximally mixed,
 $\rho^A=\rho^B={\mathbbm 1}/2$,
 so that both Bloch vectors vanish, 
 $\tau_A=\tau_B=0$,
 the correlation matrix $R$ represents a 
 positive density matrix $\rho^{AB}$ 
 if the vector $\vec v$ of three real singular values of $R$
  belongs to the regular tetrahedron  inscribed in the cube
    $[-1,1]^3$  -- see  \cite{BH300}.
  In the general case of larger dimensions and  non-zero Bloch vectors
  the conditions for the correlation matrix $R$ to assure positivity of
  the state $\rho^{AB}$ are not easy to provide in an  analytical form,
  so one has to rely on numerical techniques.
  
  \medskip
   
The Fano form   (\ref{Fano_n})
of a bipartite state is useful to describe the set
 $\Gamma^Q$ of couplings in the 
quantum transport  problem.
As the vectors $a$ and $b$ of length $N^2-1$
appearing in the matrix (\ref{Fano2}) 
represent  both partial traces, we need to fix them 
by the Bloch vectors  $\tau_A$ and $\tau_B$ 
of the analyzed states $\rho^A$ and $\rho^B$.
Thus, the minimization in Eq.\ \eqref{cost} in the main body of the article 
 is taken over the set of correlation matrices $R$
 for which the density matrix $\rho^{AB}$
 determined by matrix   (\ref{Fano2}) is positive semidefinite,
\begin{equation}
   \Gamma^Q(\rho^A, \rho^B) =
   \{  \rho^{AB}(\tau_A, \tau_B, R):   \rho^{AB}\ge 0 \}.
    \label{coupling_q}
      \end{equation}

We recall here some properties of the set of couplings
and the extremal points of this set analyzed in \cite{Ru04,Pa05,LPW14}:
 
 \medskip

  For any two states  
  $\rho^A$ and $\rho^B$ of dimension $N$ 
   the set   $ \Gamma^q (\rho^A, \rho^B)$ 
     of admissible couplings:
  
  a)  includes the product state, $\rho^A\otimes \rho^B  \in \Gamma^q (\rho^A, \rho^B)$.
   
   b) forms a convex subset of the set  $\Omega_{N^2}$ of all bipartite states. 
      If  $\rho^{AB}(\tau_A, \tau_B, R_1)  \in \Gamma^q$ 
    and $\rho^{AB}(\tau_A, \tau_B, R_2)  \in \Gamma^q$,
      then any convex combination thereof forms a density matrix, 
        $x\rho^{AB}(\tau_A, \tau_B, R_1)  + (1-x)
            \rho^{AB}(\tau_A, \tau_B, R_2)  \in \Gamma^q$ for $x\in [0,1]$,
            as any convex combination of two positive matrices is positive.
           
  c) contains a state of rank one (a projector onto a pure state) iff both arguments 
      have the same spectrum:
       $|\psi\rangle \langle \psi| \in  \Gamma^q (\rho^A, \rho^B)  
     \Leftrightarrow  {\rm Eig}(\rho^A)={\rm Eig}(\rho^B)$.  
         This statement holds  as for any pure state in a bipartite system
             both reduced density matrices (obtained  by partial traces)  
             have the same spectrum, so that
             they are unitarily similar,   $\rho^B=U\rho^A U^{\dagger}$.
             The spectrum of reduced states determines the Schmidt vector \cite{BZ17}
             of the bipartite pure state $|\psi\rangle$.

 \medskip        
  
  As a minimum of a linear function over a convex set is attained at its boundary,
  due to item c) above
  the extreme of the quantum transport  problem 
      in the  case of two states with different spectra 
   can be achieved for a  bipartite state $\rho^{AB}$
       of rank $2,3,\dots, N^2-1$.
The question concerning the relation between the spectrum of 
a  bipartite state $\rho^{AB}$
and  the spectra of its partial traces,  $\rho^A$ and $\rho^B$,
is known as the {\sl quantum marginal problem}.
 In the simplest case of two states of size $N=2$
 this problem was solved by Bravyi \cite{Br04},
 while a general theory providing the solution for larger dimensions
 was developed by Klyachko \cite{Kl04}.

\section{Transport problem and quantum operations}
   
A completely positive, trace preserving linear map  $\Phi$ acting on the set $\Omega_N$
of quantum states is called a {\sl quantum operation} or {\sl quantum channel}.
Its action on a given state $\rho$
can be conveniently written in Kraus form,
$\rho'=\Phi(\rho)=\sum_{j=1}^r K_j \rho K_j^{\dagger}$.
The number $r$ of Kraus operators $K_j$ is arbitrary, 
but to ensure the trace preserving condition
they must satisfy the identity resolution $\sum_{j=1}^r K_j^{\dagger} K_j=
{\mathbbm 1}$ -- see \cite{BZ17}. 
  
 Making use of  the  Bloch representation   (\ref{Bloch_n}),
 let us represent the initial state $\rho$ by the Bloch vector $\tau$ 
 and its image $\rho'$ by the transformed vector $\tau'$.
 In this way one can rewrite any 
 quantum operation as a linear  action on Bloch vectors,
\begin{equation}
\tau' =   Q \tau + \kappa, 
\label{Bloch2}
\end{equation} 
 where the real distortion matrix $Q$ has dimension $N^2-1$, and $\kappa$
 is a translation vector of the same length, which vanishes for unital maps.
Hence, the superoperator $\Phi$ can be represented 
 by an asymmetric  real matrix of order $N^2$,
\begin{equation}
\Phi \; =\; \left[
\begin{array}{ll}
  1 & 0 \\
{\vec \kappa}&{ Q}
\end{array}
\right] .
\label{super2}
\end{equation}
The above form, also called the {\sl Liouville  representation} of a map 
\cite{KR01, KSRJO14}, is convenient 
for spectral analysis: the spectrum of the superoperator $\Phi$
consists of the leading Frobenius--Perron  eigenvalue, 
$\lambda_1=1$, and the $N^2-1$ eigenvalues of
the real matrix $Q$, which can be complex.

Apparent similarity  between the form (\ref{super2}) of an arbitrary
operation  $\Phi$ and the matrix  (\ref{Fano2}) is not accidential,
as it is a consequence of representing the Jamio{\l}kowski--Choi state
$J=(\Phi \otimes {\mathbbm 1}) |\phi_+\rangle \langle \phi_+|$
belonging to the extended space of size $N \times N$ 
in Fano form.
Here $|\phi_+\rangle = \frac{1}{\sqrt{N}} \sum_{j=1}^{N} |j,j\rangle$
denotes the maximally entangled Bell state \cite{BZ17}.
 Note that the vector $a$ in  (\ref{Fano2})
vanishes due to the trace-preserving condition.

Returning now to the transport problem and the set
   $\Gamma^Q (\rho^A, \rho^B)$
   of admissible couplings   (\ref{coupling_q}),
   we see   that in general the Fano form (\ref{Fano2})
  {\sl  does not} describe quantum operations which send the initial state
    $\rho^A$ to the final state $\rho^B$.
    In fact, for an arbitrary initial Bloch vector
        $a=\tau_A \ne 0$,
        the corresponding transformation $\Phi$ is not trace preserving.
    However,  in the particular case
    of an initial state which is maximally mixed,  $\rho^A={\mathbbm 1}/N$,
     the corresponding Bloch vector  vanishes, $\tau_A=0$,
    and the final Bloch vector,
    $\tau_B=Q \tau_A +  b$,
      indeed represents the final state $\rho^B$ with the Bloch vector $\tau_B= b$.
     In such a case  optimization over the set of all admissible couplings 
       $
       \Gamma^Q$ can be considered as optimization over the set 
          of quantum operations,  parametrized by the distortion matrix $Q$,
           which transforms the initial maximally mixed state,
           $\rho^A={\mathbbm 1}/N$, into the final state  $\rho^B$,
           as expected in the transport problem. 
        
     The dual condition, $b=0$, corresponds to unital maps which preserve identity,
     so bistochastic  operations,  $a=b=0$, are   
      represented by Choi matrices with both states maximally mixed.
     Extremal points of the set of  these tracial states were analyzed
       by Ohno  \cite {Oh10}. 
       
       Note that the cost matrix $C^Q$ of order 4, defined in Eq.\ \eqref{cost_alpha} in the main body of the article, forms the Choi matrix corresponding to the rotation $\Phi_y$ with respect to the $y$ axis,
\begin{align}
C^Q = (\Phi_y \otimes \bone) \ket{\phi_+} \bra{\phi_+} =
(\sigma_2 \otimes \bone) \ket{\phi_+} \bra{\phi_+} (\sigma_2 \otimes \bone),
\end{align}
equivalent to the projector onto the singlet state $\ket{\psi_-}$.  

\medskip
   
The problem of quantum optimal transport  was recently related to the
question of determining  the distinguished 
quantum channel that,  for a given input state,
produces a prescribed output state \cite{DPT19,Du20a}.

\end{document}